\documentclass[aps,pra,twocolumn,superscriptaddress]{revtex4-2}

\usepackage[dvipsnames]{xcolor}

\definecolor{darkred}{rgb}{0.6,0.05,0.05}
\definecolor{darkgreen}{rgb}{0.05,0.6,0.05}
\definecolor{darkappa_blue}{rgb}{0.05,0.05,0.6}
\usepackage[colorlinks]{hyperref}
\hypersetup{colorlinks,
linkcolor=darkred,
filecolor=darkgreen,
urlcolor=darkappa_blue,
citecolor=darkgreen}
\usepackage{tabularx}
\usepackage{array}

\usepackage{tikz}
\usepackage{mathtools}
\usepackage{amsmath}
\allowdisplaybreaks
\usepackage{ bbold }
\usepackage{amssymb}
\usepackage[normalem]{ulem}
\usepackage{amsthm}
\usepackage{xfrac}
\usepackage{dsfont}
\usepackage{siunitx}
\usepackage{physics}
\usepackage{graphicx}
\usepackage{hyperref}
\usepackage[capitalise]{cleveref}
\usepackage{dcolumn}
\usepackage{bm}
\usepackage{layouts}
\usepackage{comment}
\usepackage{soul}

\newcommand{\rmd}{{\rm d}}
\newcommand{\rmi}{{\rm i}}
\newcommand{\LL}{{\mathcal{L}}}

\begin{document}

\author{Filippo Ferrari}
\email{filippo.ferrari@epfl.ch}
\affiliation{Laboratory of Theoretical Physics of Nanosystems (LTPN), Institute of Physics, \'{E}cole Polytechnique F\'{e}d\'{e}rale de Lausanne (EPFL), 1015 Lausanne, Switzerland}
\affiliation{Center for Quantum Science and Engineering, \\ \'{E}cole Polytechnique F\'{e}d\'{e}rale de Lausanne (EPFL), CH-1015 Lausanne, Switzerland}
\author{Joachim Cohen}
\affiliation{Alice \& Bob, 53 boulevard du G\'en\'eral Martial Valin, 75015 Paris, France}
\author{Vincenzo Savona}
\affiliation{Laboratory of Theoretical Physics of Nanosystems (LTPN), Institute of Physics, \'{E}cole Polytechnique F\'{e}d\'{e}rale de Lausanne (EPFL), 1015 Lausanne, Switzerland}
\affiliation{Center for Quantum Science and Engineering, \\ \'{E}cole Polytechnique F\'{e}d\'{e}rale de Lausanne (EPFL), CH-1015 Lausanne, Switzerland}
\author{Fabrizio Minganti}
\email{fabrizio.minganti@alice-bob.com}
\affiliation{Alice \& Bob, 53 boulevard du G\'en\'eral Martial Valin, 75015 Paris, France}

\title{Bit flips, saturation, and quantum chaos in dissipative cat qubits}

\date{\today}

\begin{abstract}
Bosonic cat qubits promise hardware-efficient quantum error correction because their logical bit-flip rate is exponentially suppressed with the photon number of the cat state.
However, several experiments report a saturation of this suppression at large photon numbers, thus limiting the achievable protection.
Combining quantum-trajectory simulations, semiclassical analysis, and Liouvillian spectral methods, we investigate the properties of bit flips in realistic dissipative cat qubits, where a memory mode hosting quantum information interacts with a dissipative buffer cavity.
We show that bit flips are dynamical processes inherently involving both the memory and buffer, and therefore cannot be captured by single-mode approximate descriptions.
We identify a reflection symmetry, resulting in a phase-locking condition at the semiclassical level and for quantum trajectories, as the main requirement for regular bit-flip dynamics. Its breakdown is the origin of the saturation, and we find that it occurs when two conditions are met.
First, the adiabatic approximation, where the state of the buffer instantaneously follows that of the memory, must not be valid, which typically happens at large photon numbers.
Second, key parameters such as the cross-Kerr interaction and dephasing must be present, leading to irregular dynamics in which memory fluctuations are amplified by the buffer during bit flips.
In this regime, we find that bit flips manifest as chaotic bursts within otherwise regular dynamics, as evidenced by both changes in the topology of quantum trajectories and in the Liouvillian spectrum and its associated eigenmodes involved in these switching events.
Finally, we verify our predictions against experimental data, highlighting the detrimental role of dissipative chaotic behavior in bosonic error-correcting codes.
\end{abstract}

\maketitle

\section{Introduction}

\subsection{Bosonic cat qubits and error protection}

Bosonic quantum systems provide a versatile platform for encoding and protecting quantum information.
Superconducting circuits~\cite{krantz_quantum_2019, blais_circuit_2021}, trapped ions~\cite{leibfried_quantum_2003, bruzewicz_trapped-ion_2019}, and photonic platforms~\cite{kok_linear_2007, flamini_photonic_2019, wang_integrated_2020} use bosonic excitations to manipulate and encode quantum information.
In particular, bosonic quantum codes leverage the Hilbert space of bosonic modes to perform quantum error correction~\cite{cochrane_macroscopically_1999, gottesman_encoding_2001, albert_performance_2018, lieu_symmetry_2020}. 
By appropriately choosing the codewords (the logical 0 and 1), quantum information can be robustly encoded within a single physical degree of freedom, providing a hardware-efficient route to fault-tolerant quantum computation by reducing the number of physical components per logical qubit \cite{leghtas_hardware-efficient_2013, mirrahimi_dynamically_2014, michael_new_2016, grimsmo_quantum_2020, putterman_hardware-efficient_2025}.

\begin{figure}
    \centering
    \includegraphics[width=0.975\linewidth]{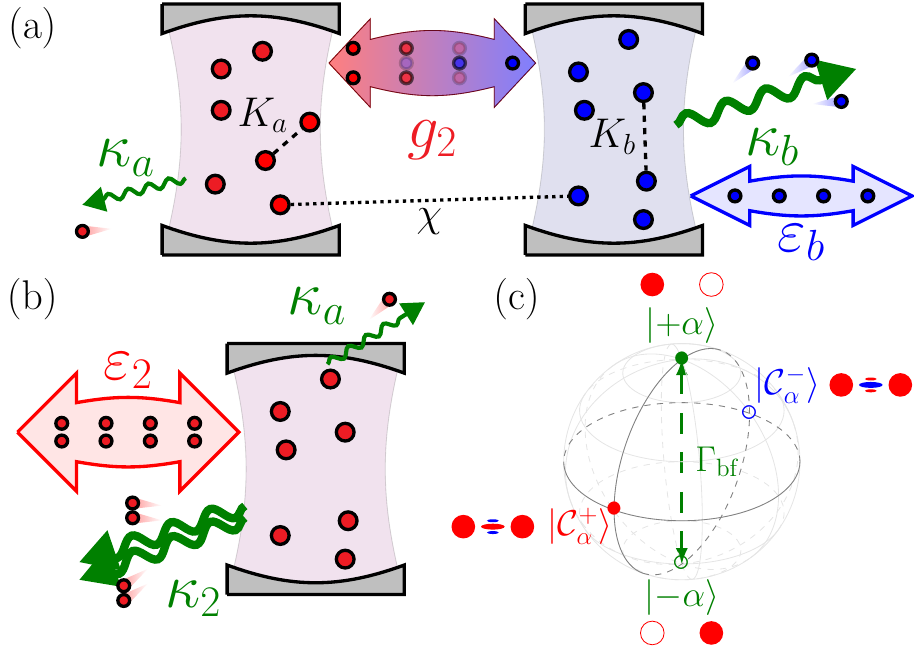}
    \caption{Pictorial description of the Hamiltonian and dissipators stabilizing cat states.
    (a) Eq.~\eqref{eqs:two_body} describes a memory (left resonator) coupled to a buffer (right resonator) via two-photon downconversion processes of strength $g_2$. 
    The buffer is driven with amplitude $\varepsilon_b$, and Kerr ($K_a$ and $K_b$) and cross-Kerr ($\chi$) nonlinearities are also present.
    Both resonators lose photon at rates $\kappa_a\ll \kappa_b$ and dephasing at rate $\kappa_\phi$ affects the $a$ mode (not shown).    
    (b) In the adiabatic limit, the evolution of the memory is captured by Eq.~\eqref{eq:single_body}. $\kappa_2$ is the two-photon dissipation rate and $\varepsilon_2$ the two-photon drive amplitude. 
    (c) The Bloch sphere defined for the single-mode system in (b) if $\kappa_a = \kappa_\phi = 0$. 
    The coherent states $\ket{\pm\alpha}$ (cat states $\ket{\mathcal{C}^\pm_\alpha}$) define the logical states $\ket{\pm_z}$ ( $\ket{\pm_x}$).
    Including photon loss and dephasing introduces bit flips at a rate $\Gamma_{\rm bf}$.
    }
    \label{fig:scheme}
\end{figure}

Schr\"odinger cat codes encode quantum information in coherent states with opposite phases \cite{cochrane_macroscopically_1999,leghtas_hardware-efficient_2013, mirrahimi_dynamically_2014}.
In superconducting circuits, cat states can be dissipatively stabilized by engineering parametric exchanges between a high-coherence memory resonator and a strongly dissipative ancillary buffer mode \cite{leghtas_hardware-efficient_2013, mirrahimi_dynamically_2014, leghtas_confining_2015, touzard_coherent_2018, lescanne_exponential_2020, reglade_quantum_2024, marquet_autoparametric_2024, putterman_preserving_2025} (see Fig.~\ref{fig:scheme} and Table~\ref{tab:previous_literature}).
We will refer to them as \textit{dissipative cats}.
In the adiabatic limit the buffer instantaneously follows the memory, whose state effectively evolves under the action of two-photon drive and dissipation.
The system is extremely constrained and analytical formulas for the steady state \cite{minganti_exact_2016,roberts_driven-dissipative2020} and dynamics \cite{carde_nonperturbative_2026} predict an exponential decrease of the bit-flip rate at the cost of a linear increase in the phase-flip one as a function of the cat size (i.e., mean photon number) \cite{mirrahimi_dynamically_2014}.
Indeed, common noise processes, such as single-photon loss or dephasing, predominantly act within the coherent-state manifold and therefore rarely induce bit flips.
This tradeoff makes cat qubits a paradigmatic example of noise-biased qubits and underpins many error correction strategies~\cite{guillaud_repetition_2019, puri_bias-preserving_2020, chamberland_building_2022, ruiz_ldpc-cat_2025}.

Beyond dissipative stabilization, Kerr photon-photon interaction together with two-photon drives also confine the system dynamics to the low-energy manifold spanned by cat states~\cite{goto_bifurcation-based_2016, goto_universal_2016, puri_engineering_2017, puri_stabilized_2019, grimm_stabilization_2020, iyama_observation_2024, hajr_high-coherence_2024}, but noise seems to ultimately limit bit-flip suppression \cite{benhayounekhadraoui2025kerrcatqubitdies}.
Various proposals for the enhancement of cats performance have been devised and implemented~\cite{schlegel_quantum_2022, hillmann_quantum_2023, rousseau2025enhancingdissipativecatqubit}, and hybrid configurations that combine Kerr and dissipative stabilization have been explored~\cite{gautier_combined_2022,ruiz_two-photon_2023,gravina_critical_2023,labaymora2025chiralcatcodeenhanced} and experimentally probed~\cite{adinolfi2025enhancingkerrcatqubitcoherence}.

\subsection{Saturation of bit-flip rate in dissipative cats}

While theoretical results predict an exponential reduction of the bit-flip error rate, experiments reveal important deviations from this scaling~\cite{lescanne_exponential_2020,berdou_one_2023, marquet_autoparametric_2024, reglade_quantum_2024,putterman_preserving_2025}. For moderate-size cats additional buffer-mediated dephasing can explain the reduction in bit-flip protection  \cite{putterman_preserving_2025} (see also Table~\ref{tab:previous_literature}).
For larger cats, however, the bit-flip rate stops decreasing with the cat size and \textit{saturates}, falling well short of the anticipated tens of minutes bit-flip times.
Single-mode models fail to even qualitatively reproduce the onset of saturation \cite{berdou_one_2023,reglade_quantum_2024,putterman_preserving_2025}, even when additional dissipation mechanisms, such as dephasing or heating, are included.

This saturation limits the viability of dissipative cats as hardware-efficient logical qubits.
Thus, understanding the microscopic origin of these bit flips, their scaling with cat size, and the mechanisms leading to saturation is essential to realize the full potential of cat codes.

\begin{table*}[t]
    \centering
    \small
    \setlength{\tabcolsep}{4pt}
    \begin{tabular}{c|c|c|c}
        & Kerr $=0$ & Kerr $\neq 0$ & $\chi \neq 0$ \\
        \hline
        $g_2 =0$
        & Impossible
        & Staircase behavior
          \cite{frattini_observation_2024,ruiz_two-photon_2023}
        & Possible by extending
          \cite{benhayounekhadraoui2025kerrcatqubitdies} \\
        (Kerr cat)
        & 
        & & for interaction with the readout
         \\
        \hline
        $g_2 \ll \kappa_b$
        & Ideal dissipative cat with
        & Hybrid cats with possible use
        & $\chi$-induced dephasing degrades \\
        (Adiabatic limit)
        & exponential protection
          \cite{mirrahimi_dynamically_2014,carde_nonperturbative_2026}
        & of detuning
          \cite{ruiz_two-photon_2023,gautier_combined_2022,gravina_critical_2023,labaymora2025chiralcatcodeenhanced,adinolfi2025enhancingkerrcatqubitcoherence}
        & bit flips
          \cite{putterman_preserving_2025} \\
        & [\textcolor{green!60!black}{This work}] & & \\
        \hline
        $g_2 \lesssim \kappa_b$
        & Reflection symmetry (phase-locked
        & Breakdown of the encoding
        & Chaotic bit flips induce saturation \\
        (Two-mode cat)
        & trajectories) preserve good bit-flip
        & due to bistability
          [\textcolor{green!60!black}{This work}]
        & [\textcolor{green!60!black}{This work}],
          explaining, e.g.,
          \cite{reglade_quantum_2024} \\
        & scaling
          [\textcolor{green!60!black}{This work}]
        & &
    \end{tabular}
    \caption{Parameter regimes, effective equations, and explanation of the relevant terms governing bit flips in the literature.}
    \label{tab:previous_literature}
\end{table*}

\subsection{Overview of the results}

In this work, we provide a framework for understanding bit flips in cat states and explain how and why the saturation of the bit-flip error rate emerges.
Key to this understanding is that bit flips are rare events, and terms that can be neglected at the steady-state level strongly affect the bit-flip dynamics.
We show that the interplay between nonlinearities--in particular cross-Kerr interaction-- dephasing, and nonadiabatic effects causes saturation. 

\begin{figure*}
    \centering
    \includegraphics[width=\linewidth]{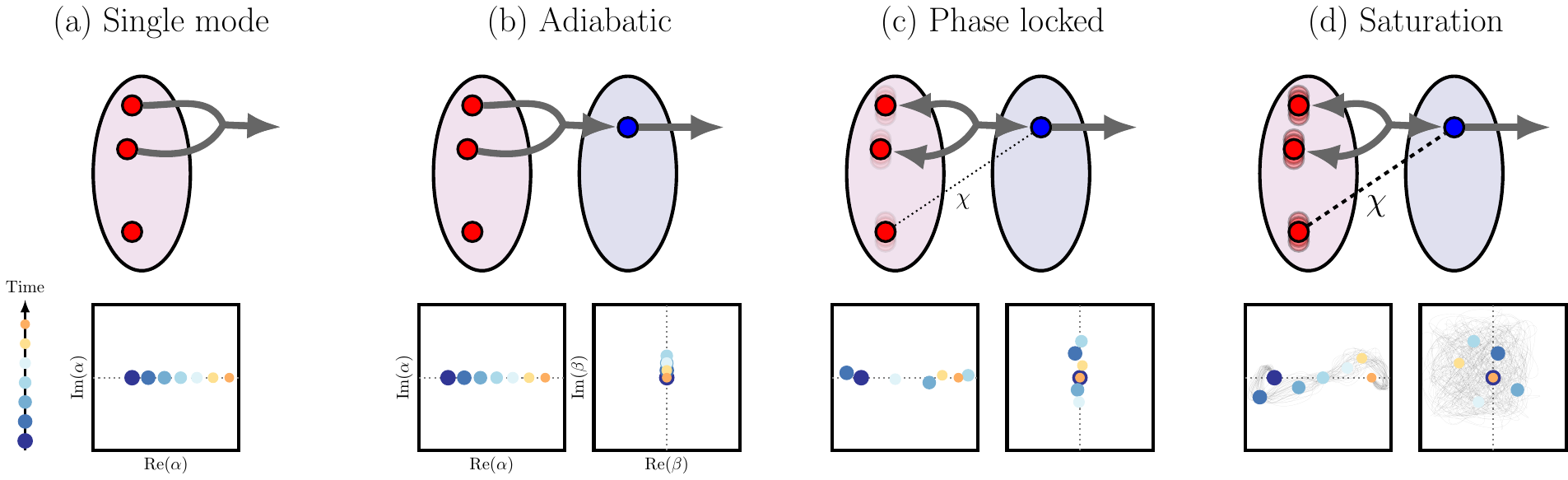}
    \caption{Behavior of the system during a bit flip. Top: Sketch of the resonators. Bottom: simplified dynamics in the phase-space.
    (a) Single-mode approximation. Bit flips are regular, as the motion is constrained on a line connecting the coherent states of opposite phase.
    (b) In the adiabatic regime, the buffer instantaneously follows the memory, leading to an effective two-photon dissipation for the memory (unidirectional arrows in the scheme). 
    The memory behaves as in the single-mode case, while the buffer only slightly moves from its stable position.
    (c) Outside of the adiabatic regime, the memory and buffer coherently exchange excitations (bidirectional arrows).
    Nonetheless, the dynamics is regular as the phase between buffer and memory is locked, up to small fluctuations.
    The buffer introduces photon-number fluctuations in the memory, only partially degrading the bit flip.
    Such a quasi phase locking persists for moderate Hamiltonian nonlinearities and memory dephasing.
    (d) As the effect of nonlinearities becomes dominant, fluctuations both in number and phase are amplified, breaking the phase locking condition.
    This ultimately leads to irregular dynamics, associated with chaotic-like behavior, and manifests as a saturation of the bit-flip scaling.
    Other nonlinearities, such as Kerr in the memory, exacerbate this effect.}
    \label{fig:scheme_adiabaticity}
\end{figure*}

Second, using quantum trajectory simulations, we show that bit flips are non-instantaneous two-mode dynamical excursions through the Hilbert space. 
Their features cannot be captured by single-mode descriptions and depend on the strength of nonlinearities and nonadiabatic effects, as sketched in Fig.~\ref{fig:scheme_adiabaticity}.
If the adiabatic approximation holds, the system switches between the two coherent in a regular fashion [Figs.~\ref{fig:scheme_adiabaticity} (a-b)]. 
As nonadiabatic effects become stronger, the buffer mediates and amplifies photon-number fluctuations of the memory [Fig.~\ref{fig:scheme_adiabaticity} (c)], while the dynamics still retains regular features. 
Such regular behavior is associated with a reflection symmetry of the Liouvillian. 
At the semiclassical level, it manifests in a phase locking condition that keeps the relative phases of memory and buffer fixed.
In quantum trajectories, the symmetry results in a \textit{quasi phase locking}: the position and momentum observables slightly fluctuate around the angles predicted by the semiclassical analysis.
Nonlinearities invalidate the reflection symmetry, and compete to break the phase locking mechanism, changing the topology of bit flips in the saturation regime [Fig.~\ref{fig:scheme_adiabaticity} (d)].

Third, we find that the saturation of bit flips is closely connected to the emergence of chaotic behavior, as revealed by a Liouvillian spectral analysis.
When quasi phase locking is lost, bit flips explore increasingly large regions of the Hilbert space, with trajectories becoming delocalized and showing features consistent with dissipative chaotic dynamics.
The emergent picture is that short and rare chaotic-like bursts, triggered by the amplification of fluctuations within an otherwise integrable dynamics, are the limiting factor for the scaling of bit flips in large cat states.
Since chaos is a universal phenomenon governed by the global structure of the operators rather than microscopic details, our analysis suggests that this mechanism is broadly relevant across dissipative cat qubits, and arguably in driven-dissipative bosonic platforms.

Finally, we validate our theoretical framework by comparing our predictions with the experimental results of Ref.~\cite{reglade_quantum_2024}.
Including cross-Kerr and dephasing--the terms we identified as necessary to trigger saturation and chaotic behavior during the bit flip--we quantitatively retrieve the experimental bit-flip times even in the otherwise unexplained saturation region.

The structure of the paper is the following.
In Sec.~\ref{sec:model_methods} we introduce the model describing dissipative cat qubits and motivate the parameter choice.
In Sec.~\ref{sec:saturation} we study the bit-flip error rate and show that saturation of bit-flip time is a consequence of nonlinear and nonadiabatic effects.
In Sec.~\ref{sec:bit_flips} we characterize bit flips using quantum trajectories, showing an increase in the complexity of the dynamics from the adiabatic regime to the nonlinear and nonadiabatic one, where saturation occurs.
In Sec.~\ref{sec:semiclassical_analysis} we identify the presence of a reflection symmetry as central element in determining regular bit flips and, by analyzing quantum trajectories and through a semiclassical treatment of the equations of motion, we show that this results in a locking of the relative phases of buffer and memory.
In Sec.~\ref{sec:dissipative_quantum_chaos} we discuss the emergence of dissipative quantum chaos in cat qubits in the regime where bit flip saturates.
In Sec.~\ref{sec:experiment} we compare our theoretical predictions with the experimental data of Ref.~\cite{reglade_quantum_2024}.
Finally, we draw our conclusions in Sec.~\ref{sec:discussion}.

\section{Model}\label{sec:model_methods}

As sketched in Fig.~\ref{fig:scheme} (a), we consider two coupled driven-dissipative nonlinear bosonic resonators described by annihilation operators $\hat{a}$ (memory) and $\hat{b}$ (buffer). 
This setup represents a typical circuit implementation of a dissipative cat qubit \cite{lescanne_exponential_2020,reglade_quantum_2024,marquet_autoparametric_2024,putterman_preserving_2025}.
The system density matrix $\hat{\rho}$ evolves according to the Lindblad master equation (see, e.g., Refs.~\cite{guillaud_quantum_2023, berdou_one_2023}):
\begin{equation}\label{eqs:two_body}
\begin{split}
    \frac{\partial\hat{\rho}}{\partial t} = & -\rmi[\hat{H}, \hat{\rho}] + \kappa_b\mathcal{D}[\hat{b}]\hat{\rho}  + \kappa_a\mathcal{D}[\hat{a}]\hat{\rho} + \kappa_\phi\mathcal{D}[\hat{a}^\dagger\hat{a}]\hat{\rho} , \\
    \hat{H} =\,& g_2\left(\hat{a}^{\dagger 2}\hat{b} + \hat{a}^2\hat{b}^\dagger\right)  + \left( \varepsilon_b \, \hat{b}^\dagger + \varepsilon_b^* \,\hat{b}\right)\\
    & \quad + \frac{K_a}{2}\hat{a}^{\dagger 2}\hat{a}^2 + \frac{K_b}{2}\hat{b}^{\dagger 2}\hat{b}^2 + \chi \hat{a}^{\dagger}\hat{a} \hat{b}^{\dagger }\hat{b}. 
\end{split}
\end{equation}
The Hamiltonian $\hat{H}$ includes a two-photon exchange interaction of strength $g_2$ and a coherent drive on the buffer mode with amplitude $|\varepsilon_b|$.
Typically, $g_2$ is tuned using driven and magnetically pumped nonlinear elements~\cite{leghtas_confining_2015,lescanne_exponential_2020}. 
Imperfections in fabrication~\cite{reglade_quantum_2024} or dynamically activated processes~\cite{carde_flux-pump-induced_2025} typically result in Kerr nonlinearities in the memory ($K_a$) and buffer ($K_b$), as well as a cross-Kerr interaction $\chi$ between the two modes.

Equation \eqref{eqs:two_body} includes the dissipators $\mathcal{D}[\hat{L}]\hat{\rho} = \hat{L} \hat{\rho} \hat{L}^\dagger - \{\hat{L}^\dagger \hat{L}, \hat{\rho}\}/2$, where $\hat{L}$ denotes a Lindblad jump operator; $\kappa_a$ ($\kappa_b$) is the photon loss rate in the memory (buffer), while $\kappa_\phi$ describes dephasing in the memory.
While $\kappa_b$ is essential to stabilize cats, $\kappa_a$ and $\kappa_\phi$ are sources of unwanted noise, inducing bit- and phase-flip errors.
Additional terms can appear in Eq.~\eqref{eqs:two_body}, e.g., detunings between the modes and the pump or driving fields.
Since we focus on regimes where dissipative stabilization dominates, we neglect such detunings throughout this work,  as done in several experiments \cite{leghtas_confining_2015,lescanne_exponential_2020, berdou_one_2023, putterman_hardware-efficient_2025}.

We introduce the Liouvillian superoperator $\LL$ and recast Eq.~\eqref{eqs:two_body} as
\begin{equation}\label{eq:liouvillian}
    \frac{\partial\hat{\rho}}{\partial t} = \LL \hat{\rho}.
\end{equation}
Since $\LL$ is linear and non-Hermitian, it admits left and right eigenoperators defined by
\begin{equation}\label{eq:liouvillian_spectrum}
    \LL\,\hat{\eta}_j = \lambda_j\,\hat{\eta}_j,\qquad\qquad
    \LL^\dagger\,\hat{\sigma}_j=\lambda_j^*\,\hat{\sigma}_j,
\end{equation}
with the biorthonormality condition $\operatorname{Tr}(\hat{\sigma}_j^\dagger\,\hat{\eta}_k) = \delta_{jk}$.
This spectral decomposition provides full access to the system dynamics, and thus to the bit- ($\Gamma_{\rm bf}$) and phase-flip ($\Gamma_{\rm pf}$) rates.
In particular, for the cases considered in this work, $\Gamma_{\rm bf}=-\lambda_1$ is the Liouvillian gap, with $\lambda_1$ being the eigenvalue with the smallest nonzero real part.
The steady state $\hat\rho_{\rm ss}$ such that $\partial \hat\rho_{\rm ss}/\partial t =0$ is instead associated with $\lambda_0 =0$.
More details on how to associate the eigenvalues of the Liouvillian to bit- and phase-flip rates in more general cases can be found in, e.g., Refs.~\cite{gravina_critical_2023, labaymora2025chiralcatcodeenhanced}.

\subsection{Single-mode approximation}

In the limit $\kappa_b \to \infty$, the buffer mode can be adiabatically eliminated, yielding an effective Lindblad master equation for the memory mode alone [see Fig.~\ref{fig:scheme} (b)]:
\begin{equation}\label{eq:single_body}
    \frac{\partial\hat{\rho}}{\partial t} = -\rmi[\hat{H}, \hat{\rho}] 
    + \kappa_\phi\mathcal{D}[\hat{a}^\dagger\hat{a}]\hat{\rho} 
    + \kappa_a\mathcal{D}[\hat{a}]\hat{\rho} 
    + \kappa_2\mathcal{D}[\hat{a}^2]\hat{\rho},
\end{equation}
with 
$\hat{H} = \varepsilon_2(\hat{a}^2 + \hat{a}^{\dagger 2})/2 
+ K_a \hat{a}^{\dagger 2}\hat{a}^2/2$.
The parameters in Eq.~\eqref{eq:single_body} are related to those in Eq.~\eqref{eqs:two_body} through 
\begin{align}\label{eqs:conversion_formulas}
    \kappa_b = \frac{4|g_2|^2}{\kappa_2},
    \qquad 
    \varepsilon_b = -\frac{\varepsilon_2 \kappa_b}{4\rmi g_2}.
\end{align}
We further introduce $\theta$, 
\begin{equation}\label{Eq:nonlinearities}
    K_a/\kappa_2 = \tan(\theta),
    \qquad 
    \theta \in [0,\pi/2],
\end{equation}
interpolates between purely dissipative ($\theta = 0$) and Kerr-stabilized regimes ($\theta = \pi/2$), as they can jointly confine the cat manifold.
In the absence of single-photon loss and dephasing ($\kappa_a=\kappa_\phi=0$), the coherent states $\ket{\alpha = \pm \rmi \sqrt{\varepsilon_2/\kappa_2}}$ satisfying $\hat{a}\ket{\alpha} = \alpha \ket{\alpha}$ are stable, as well as their quantum superpositions.
The logical basis is then defined in the limit of large $\alpha$ as $\ket{\pm_z} = \ket{\pm\alpha}$, $\ket{\pm_x} = \ket{\mathcal{C}^\pm_\alpha} \propto \ket{\alpha} \pm  \ket{-\alpha}$, and $\ket{\pm_y} \propto (\ket{\alpha} \pm \rmi \ket{-\alpha})$, as shown in Fig.~\ref{fig:scheme} (c).

This single-mode description provides an effective model commonly used to analyze dissipative cat qubits,
and it serves as a reference against which the full two-mode dynamics will be compared in this work.

\subsection{Adiabaticity, nonlinearity, and parameter values}\label{sec:parameters}

Three ingredients lead to an increase in the complexity of bit-flip dynamics and modify the scaling of bit-flip errors, as summarized in Fig.~\ref{fig:scheme_adiabaticity}: 
(i) The lack of adiabaticity. As the buffer mode becomes incapable of dissipating photons faster than they are transferred from the memory, the prediction of the single-mode system is unreliable. 
Nonetheless, the dynamics remain regular.
(ii) Nonlinear terms, especially cross-Kerr interaction. 
Outside the adiabatic limit, where they have only marginal effects, nonlinearities amplify fluctuations between buffer and memory.
(iii) Dephasing. In the configurations we analyzed, it usually provide an initial fluctuation of the memory field then amplified by nonlinearities and nonadiabatic exchanges.

To investigate the role of adiabaticity, we analyze the system for varying values of $g_2$ and $\kappa_b$.
We choose to keep $\kappa_2$ fixed, so that, in the limit $g_2|\alpha| \ll \kappa_b$ where adiabaticity holds, all choices reduce to the same single-mode model described by Eqs.~\eqref{eq:single_body}.
Concerning nonlinearities, usually, the buffer mode has large zero-point phase fluctuations in the nonlinear element (e.g., an asymmetrically threaded SQUID \cite{lescanne_exponential_2020}), making $K_b$ and $\chi$ non-negligible.
We then compare the linear case corresponding to $K_b=\chi=0$ with cases having realistic values  $K_b/2\pi=1\,$MHz and $\chi/2\pi=0.5\,$MHz. 
When investigating different values of Kerr nonlinearities in the memory, instead, we fix the total effective nonlinearity of the reduced model, $
\Xi = \sqrt{\kappa_2^2 + K_a^2}$, which sets the overall confinement scale and thus acts as the fundamental rate of the problem \footnote{ 
Indeed, in the adiabatic limit and for $\kappa_a = \kappa_\phi = 0$, this choice sets the total number of photons in the memory for a given $\varepsilon_2$ \cite{minganti_exact_2016,roberts_driven-dissipative2020}}.
In what follows, we set $\Xi/2\pi = 1$MHz.
Finally, we will assume small but finite values of $\kappa_a/2\pi= 10\,$kHz and $\kappa_\phi/2\pi \in[0,  10]\,$kHz.

\begin{figure*}[t]
\centering
    \begin{minipage}[c]{0.15\textwidth}
        \centering
        \includegraphics[width=\textwidth, trim={0.3cm 0 0.2cm 0.95cm}]{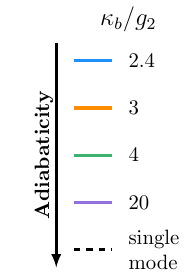} 
    \end{minipage}
    \begin{minipage}[c]{0.8\textwidth}
        \centering
        \includegraphics[width=\textwidth]{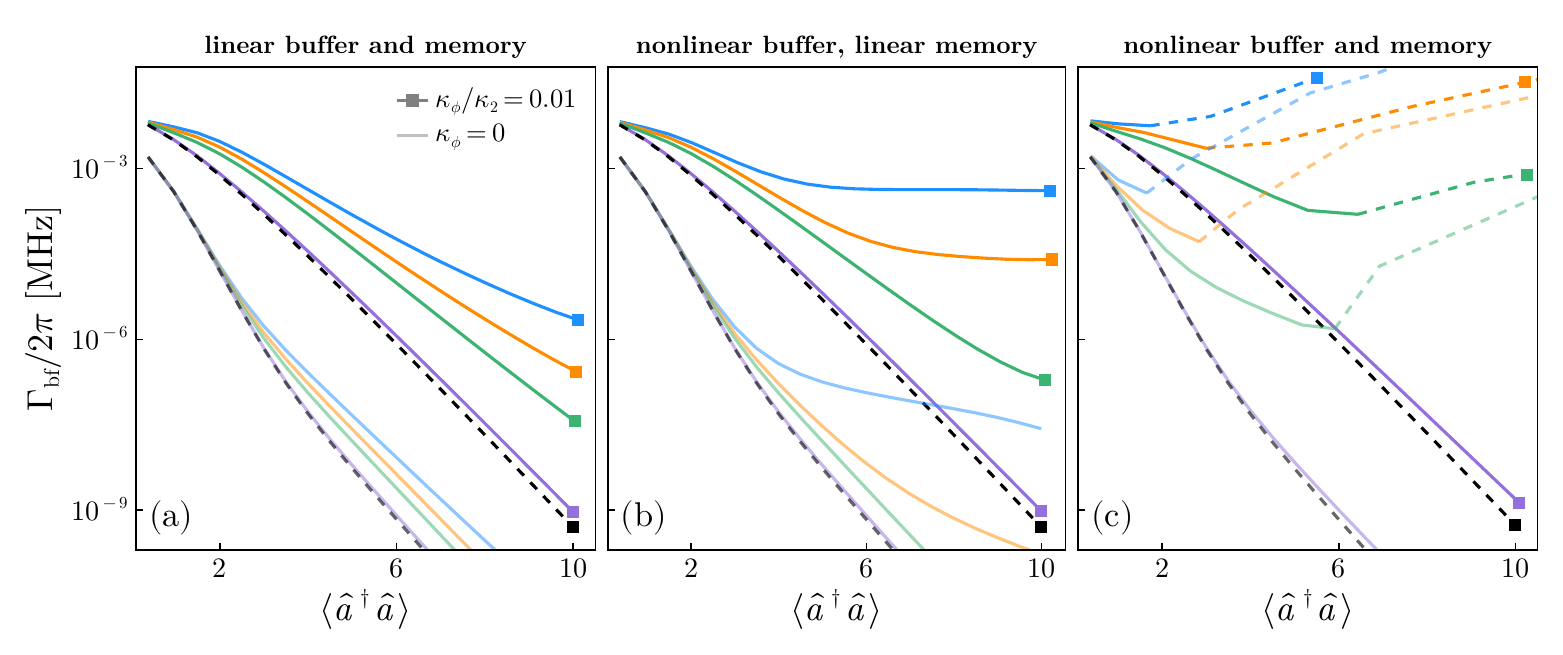}
    \end{minipage}
\caption{Bit-flip error rate $\Gamma_{\rm bf}$ as a function of the photon-number in the memory.
The total amount of nonlinearity of the system has been fixed to $\Xi/2\pi =1$MHz (see the discussion in Sec.~\ref{sec:parameters}).
We consider dephasing rates $\kappa_\phi/2\pi = 10\,\textrm{kHz}$ (darker lines ending with a square dot) and $\kappa_\phi=0$ (lighter lines)
We compare four regimes: (i) nonadiabatic ($\kappa_b/2\pi = 1.44$MHz, $g_2/2\pi =0.6$MHz, blue curves); (ii) barely nonadiabatic ($\kappa_b/2\pi = 2.25$MHz, $g_2/2\pi =0.75$MHz, orange curves); (iii) almost adiabatic ($\kappa_b/2\pi = 4$MHz, $g_2/2\pi =1$MHz, green curves); (iv) fully adiabatic ($\kappa_b/2\pi = 100$MHz $g_2/2\pi =5$MHz, purple curves).
Using Eq.~\eqref{eqs:conversion_formulas}, in the single-mode limit all these correspond to $\kappa_2/2\pi = 1$MHz.
(a) $\Gamma_{\rm bf}$ in the linear regime ($K_a=K_b=\chi=0$) for different values of the parametric downconversion rate, spanning the regimes from noadiabatic to fully-adiabatic.
The black-dashed line indicate the result for the idealized single-mode cat.
(b) Same as in (a) but in the nonlinear regime with $K_b/2\pi=1\,$MHz and $\chi/2\pi=0.5\,$MHz.
(c) Same as in (b), but in a system with small nonlinearity in the memory [$\theta=\pi/32$, see Eq.~\eqref{Eq:nonlinearities}].
The memory single-photon loss rate is fixed to $\kappa_a/2\pi = 10\,$kHz.
}
\label{fig:liouvillian_gap_dissipative}
\end{figure*}

While these parameters are in line with those reported in experiments \cite{reglade_quantum_2024, putterman_hardware-efficient_2025}, the behavior we observe is expected to be independent of this precise choice, as we argue in Sec.~\ref{sec:semiclassical_analysis} using semiclassical analysis and further demonstrate in Sec.~\ref{sec:dissipative_quantum_chaos} showing that saturation is consistent with chaotic features.
The numerical results presented in this article have been obtained using \texttt{QuantumToolbox.jl} \cite{mercurio_quantum_2025}, \texttt{Qutip} \cite{johansson_qutip_2012} and \texttt{Dynamiqs} \cite{guilmin2025dynamiqs}.

\section{Scaling of the bit-flip error rate}\label{sec:saturation}

The regimes summarized in Fig.~\ref{fig:scheme_adiabaticity} are analyzed quantitatively in Fig.~\ref{fig:liouvillian_gap_dissipative} through the scaling of the bit-flip rate $\Gamma_{\rm bf}$ as a function of the number of photons in the memory at the steady state $\langle \hat{a}^\dagger \hat{a} \rangle = \operatorname{Tr}[\hat\rho_{\rm ss} \hat{a}^\dagger \hat{a}]$.
This quantity is a key figure of merit for cat qubits, as the protection mechanism relies on an exponential suppression of bit-flip events with increasing cat size. 
In particular, we consider for both $\kappa_\phi=0$ and $\kappa_\phi \neq 0$: 
$K_a = \chi = K_b =0$ [Fig.~\ref{fig:liouvillian_gap_dissipative} (a)]; $\chi,\, K_b \neq 0$ but $K_a =0$ [Fig.~\ref{fig:liouvillian_gap_dissipative} (b)]; $\chi, K_a, \, K_b \neq 0$ [Fig.~\ref{fig:liouvillian_gap_dissipative} (c)].

As a baseline, we use the single-mode model described by Eq.~\eqref{eq:single_body} (black dashed curves). 
In this approximation the bit-flip rate always decreases exponentially with the photon number. 
Moderate Kerr nonlinearities in the memory do not qualitatively modify this scaling, and dephasing only alter the prefactor of such exponential scaling \cite{gautier_combined_2022} (see also Appendix~\ref{sec:appendix_single_mode_lergedeph}).
The single-mode model in Eq.~\eqref{eq:single_body} is not capable of retrieving a saturation of $\Gamma_{\rm bf}$.
This remains true also when considering larger values of dephasing as well as buffer-induced effective fluctuations mediated by cross-Kerr interaction \cite{putterman_preserving_2025}, as we demonstrate in Appendix~\ref{sec:appendix_single_mode}.

We now turn to the full two-mode model of Eq.~\eqref{eqs:two_body}. 
To evaluate the effects of deviations from adiabaticity, we consider $g_2/2\pi \in [0.6,5]\,$MHz and $\kappa_b/2\pi \in [1.44,100]\,$MHz.
According to the ratio $g_2/\kappa_b$, we focus on four regimes: (i) nonadiabatic (blue curves); (ii) barely nonadiabatic (orange curves); (iii) almost adiabatic (green curves); (iv) fully adiabatic (purple curves).
Despite leading to the same single-mode model in the adiabatic limit (see Sec.~\ref{sec:parameters}), we observe that, for all the studied parameters, the fully adiabatic case closely matches the prediction of the single-mode cat, deviations from adiabaticity always lead to higher values of $\Gamma_{\rm bf}$, and the presence of dephasing exacerbates the differences between single- and two-mode simulations.

We now analyze in more detail Fig.~\ref{fig:liouvillian_gap_dissipative} (a), where nonlinearities are absent ($K_a=K_b=\chi=0$). 
Deviations from adiabaticity increase $\Gamma_{\rm bf}$ but do not destroy the exponential scaling with photon number, and even far from the adiabatic limit the system retains the essential protection mechanism of the cat encoding.
This is especially true if $\kappa_\phi = 0$.
We explain this robustness in Sec.~\ref{sec:semiclassical_analysis}.

The case with $K_a=0$ and $K_b, 
\, \chi \neq 0$ is reported in Fig.~\ref{fig:liouvillian_gap_dissipative} (b).
If $\kappa_{\phi} =0$, even if the suppression of bit flips becomes sub-optimal, the system still maintains a quasi-exponential scaling except for the non-adiabatic case.
Here, we observe a severe deviation from the exponential scaling, even if the bit flip error rate reaches remarkably small values.

The situation changes \textit{qualitatively} once \textit{both} cross-Kerr nonlinearities and dephasing are present ($\chi \neq 0$, $\kappa_\phi \neq 0$).
We observe a \textit{saturation} of the bit-flip error rate with increasing photon number in both nonadiabatic curves, and we observe deviation from an exponential scaling also in the almost adiabatic case.
Although in this analysis we assumed that both $K_b$ and $\chi$ are nonzero, in Appendix~\ref{sec:appendix_two_mode_role} qualitatively similar results are obtained when setting $K_b=0$ and $
\chi \neq 0$, while no saturation emerges if $\chi =0$.
\textit{We conclude that cross-Kerr is the key nonlinearity triggering saturation}.

Finally, in Fig.~\ref{fig:liouvillian_gap_dissipative} (c) we investigate the effect of a small memory Kerr nonlinearity $K_a$ ($\theta = \pi/32$, $K_a/\kappa_2=0.1$).
$\Gamma_{\rm bf}$ abruptly increases once the photon number exceeds a critical values where optical bistability occurs \cite{drummond_quantum_1980, drummond_quantum_1981, bartolo_exact_2016}, and the logical qubit is no longer well defined (we mark this region using dotted lines).
We discuss this phenomenon and the associated states more in detail in Appendix~\ref{sec:appendix_kerr_memory_positive}.
These results indicate that even small values of memory self-Kerr can lead to a catastrophic degradation of cat-qubit performance.

A key conceptual result emerges from this analysis: the saturation of bit flip error can be self-consistently explained within a two-mode model description as a consequence of nonlinear and nonadiabatic dynamics in the presence of dephasing.
Neither ingredient alone is sufficient to reproduce the experimentally observed behavior.

\section{Anatomy of bit flips}
\label{sec:bit_flips}

We have shown that nonadiabaticity and nonlinearity degrade the bit-flip error rate and induce saturation, but not \textit{why} these terms play such a central role.
To pinpoint the issue, consider the cross-Kerr interaction.
At the steady state, the photon number $n_b^{\rm ss} = \operatorname{Tr}[ \hat\rho_{\rm ss} \hat{b}^\dagger \hat{b}]\ll 1$ for all case considered in Figs.~\ref{fig:liouvillian_gap_dissipative} (a) and (b). 
Therefore, one may think the effect of $\chi$ as negligible since
\begin{equation}\label{Eq:approximation_cross_Kerr}
    \chi \hat{a}^\dagger \hat{a} \hat{b}^\dagger \hat{b} \simeq \chi n_b^{\rm ss} \hat{a}^\dagger \hat{a}  \simeq 0   
\end{equation}
While Eq.~\eqref{Eq:approximation_cross_Kerr} allows computing steady-state properties, it leads to ambiguous predictions for the bit-flip error rate.
To unveil the dynamical features of bit flips, and understanding the mechanisms behind the saturation of $\Gamma_{\rm bf}$, we resort to quantum trajectories.

\begin{figure}[t]
\centering
\includegraphics[width=0.48 \textwidth]{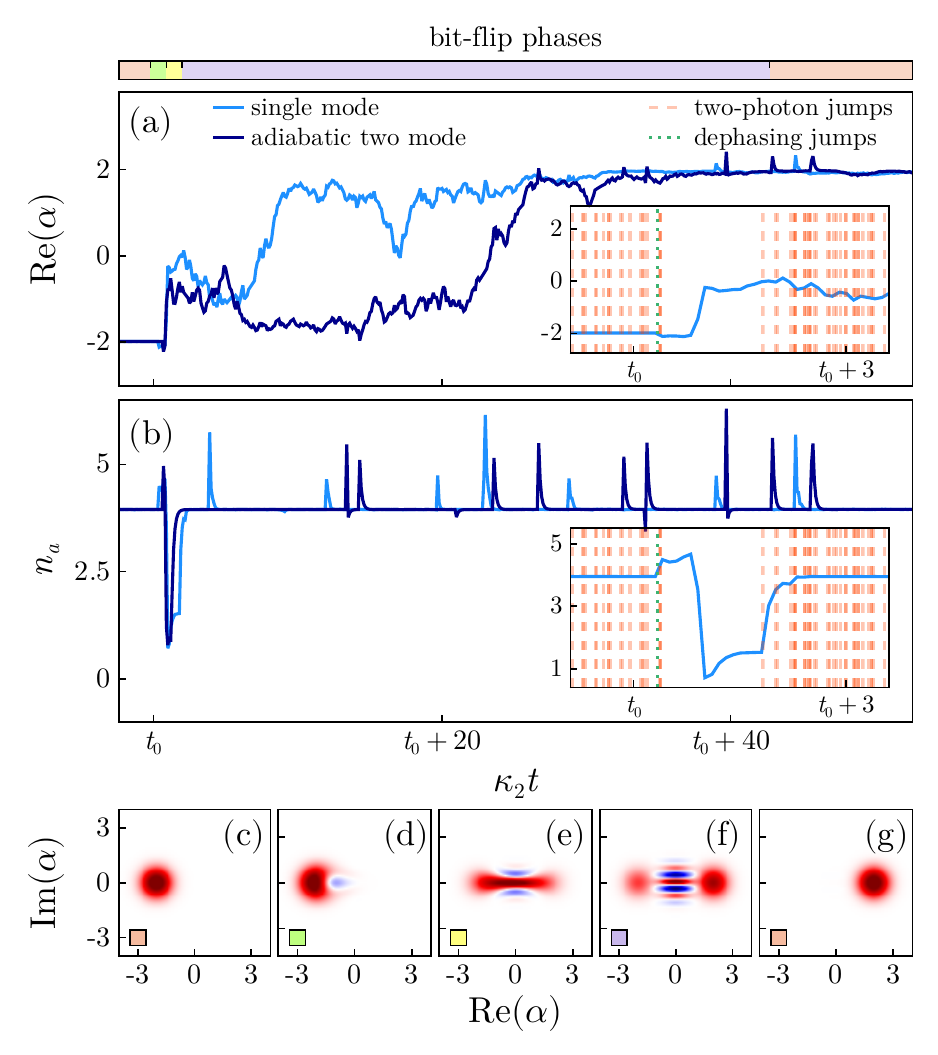}\vspace{0.02em}
\caption{Quantum trajectory $\ket{\Psi(t)}$ across a bit flip in the single-mode model (light-blue curves) and adiabatic two-mode system (dark-blue curves).
Expectation value of (a) the bosonic field $\alpha=\expval{\Psi(t)|\hat{a}|\Psi(t)}$ and (b) photon number $n_a = \langle\Psi(t)|\hat{a}^\dagger\hat{a}|\Psi(t)\rangle$.
The insets show a zoom on the initial time of the bit flip event for the single-mode model;
the red (green) vertical lines indicate two-photon (dephasing) jumps.
Here $t_0$ indicates the minimum of the photon number across the bit-flip.
The colorbar distinguishes four different phases across the bit flip: coherent phase (sienna), fluctuation phase (yellow), vacuum phase (green) and cat-like phase (purple).
(c-g) Wigner functions of $\ket{\Psi(t)}$ for the single-mode model during the phases:
(c) Coherent, preceding the bit-flip. 
(d) Fluctuation, in correspondence of the dephasing event. 
(e) Vacuum, coinciding with the absence of two-photon jumps. 
(f) Cat-like, where two-photon jumps stabilize again the cat. 
(g) Coherent, following the bit-flip.
The parameters are $\kappa_2/2\pi=1\,\textrm{MHz}$, $K_a=0$ (i.e., $\theta=\pi/2$), $\varepsilon_2/\kappa_2=4$, $\kappa_a/2\pi = \kappa_\phi/2\pi=10$ kHz.
For the two-mode system we have $g_2/2\pi = 5$ MHz, $\kappa_b/2\pi= 100$ MHz. 
We fix $\varepsilon_2$ and $\varepsilon_b$ so that the memory hosts $4$ photons in the steady state.
The $\beta$-dyne trajectory was simulated with $\beta=5$ (c.f. Appendix \ref{sec:Quantum_trajectories}).
}
\label{fig:single_mode_traj}
\end{figure}

The Lindblad master equation and the Liouvillian $\LL$ describe the ensemble-averaged dynamics of the system, but they admit stochastic unravelings in terms of \emph{quantum trajectories} (see, e.g., Refs.~\cite{carmichael_open_1993,haroche_exploring_2006,breuer_theory_2007,daley_quantum_2014} and Appendix~\ref{sec:Quantum_trajectories} for a short review).
We consider that each jump operator $\hat{L}$ appearing in a dissipator $\mathcal{D}[\hat{L}]$ results in a generalized quantum measurement performed by monitoring the exchanges between the system and the environment.
The system is described by a stochastic wave function $\ket{\Psi(t)}$, called a quantum trajectory, and evolves according to a stochastic Schr\"odinger equation.

Different measurement schemes correspond to different unravelings, but the average over many quantum trajectories always recovers $\hat\rho(t)$ in Eq.~\eqref{eq:liouvillian}.
We will assume that the buffer emission (or two-photon dissipation events in a single-mode case) and the dephasing events of the $a$ mode are measured by perfect quantum jump detectors, thus resulting in an exact counting of these events.
Single-photon jumps are instead monitored through $\beta$-dyne (i.e., homodyne-like) detection trajectories \cite{carmichael_open_1993}. 
The reason for this choice is discussed in Appendix~\ref{sec:app_hom_traj}.
The various regimes analyzed below are summarized in Fig.~\ref{fig:scheme_adiabaticity}.

\subsection{Single-mode and adiabatic two-mode cats}
\label{sec:regular_bit_flips}

\begin{figure}[t]
\centering
\includegraphics[width=0.48 \textwidth]{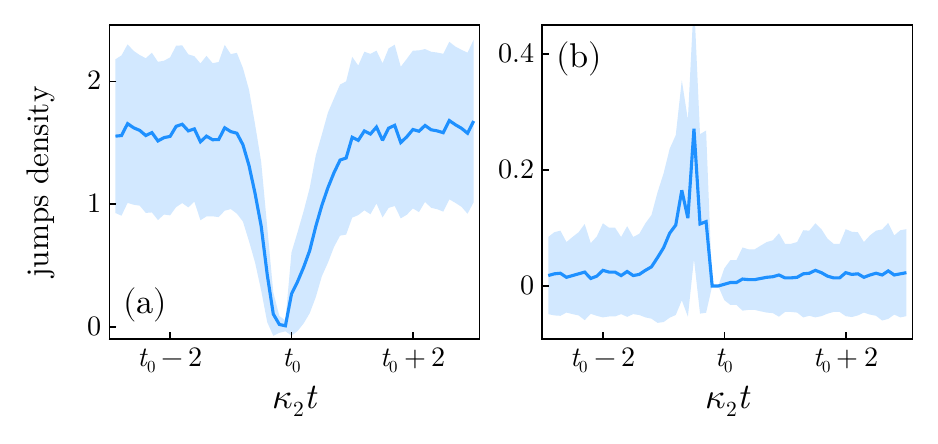}\vspace{0.01em}
\caption{
Average occurrence of quantum jumps across bit-flip events for the single-mode system.
The solid lines represent the mean density while the shaded-regions quantify the standard deviation of:
(a)  2-photon jumps ($\hat{a}^2$);
(b) dephasing jumps ($\hat{a}^\dagger\hat{a}$).
Data are collected over $10^3$ trajectories exhibiting a bit-flip, and $t_0$ indicates the time when the minimum of the photon number is reached, that we take as a reference time to synchronize different quantum trajectories.
Parameters as in Fig.~\ref{fig:single_mode_traj}.
}
\label{fig:jump_statistics}
\end{figure}

If $\kappa_a = \kappa_\phi = 0$, a coherent state $\ket{\Psi(t)} = \ket{\alpha}$ does not evolve because both the action of
two-photon jumps $\hat{a}^2$ and of the non-Hermitian Hamitlonian $\hat{H}_{\rm nh} = \hat{H} - \rmi\kappa_2 \hat{a}^{\dagger 2} \hat{a}^2/2$ leave the coherent state unaffected. 
The same holds for any superposition of $\ket{\alpha}$ and $\ket{-\alpha}$, making the logical manifold stable and immune to intra-manifold dynamics (bit- and phase-flips).

In the presence of photon loss and dephasing, the system displays bit-flip events.
Figure~\ref{fig:single_mode_traj} shows a representative trajectory exhibiting such an event.
We display the amplitude $\alpha = \langle\Psi(t)| \hat{a} |\Psi(t)\rangle$ and photon number $n_a = \langle \Psi(t)| \hat{a}^\dagger \hat{a} | \Psi(t) \rangle$, together with Wigner functions at key points of the switching process in Figs.~\ref{fig:single_mode_traj} (a), (b), and (c-g), respectively.
We observe that bit flips follow a structured temporal sequence composed of four distinct phases:
\begin{itemize}
\item \textbf{Coherent phase}:  
The system resides in a stable coherent state $\ket{\alpha}$ or $\ket{-\alpha}$.
\item \textbf{Fluctuation phase}:  
A perturbation, typically a dephasing event for the parameters considered here, displaces the state away from the logical manifold and initiates a transient fluctuation.
\item \textbf{Vacuum phase}:  
For a short but sizable time, two-photon jumps do not occur.
In their absence, the state rapidly loses photons and flows toward a squeezed-like vacuum, corresponding to the minimal-loss state of the non-Hermitian Hamiltonian. 
This rare absence of stabilizing two-photon emissions is the root cause of the bit flip.
\item \textbf{Cat-like phase}:  
Two-photon jumps resume and  confine the state back into the logical manifold.
The wavefunction temporarily develops Wigner negativities (interference fringes), but the effect of single-photon losses is to gradually suppress them within a typical time of order $1/[\kappa_a \langle\hat{a}^\dagger \hat{a}\rangle]$. 
\item \textbf{Coherent phase}: the system has collapsed either into $\ket{\alpha}$ or $\ket{-\alpha}$ with equal probability.
\end{itemize}
We confirm this structure of the bit flip performing a statistical analysis of $10^3$ trajectories in Fig.~\ref{fig:jump_statistics}. 
Indeed, bit flips require a suppression of two-photon jumps [Fig.~\ref{fig:jump_statistics} (a)], and the error is often seeded by a dephasing jump [Fig.~\ref{fig:jump_statistics} (b)].
As shown in Appendix~\ref{sec:appendix_single_mode_nodeph}, this structure--and in particular the absence of stabilizing two-photon jumps during the switching event--also characterizes the minimal setup where $\kappa_\phi=0$ but $\kappa_a \neq 0$.

We now extend the analysis to the full two-mode system of Eq.~\eqref{eqs:two_body}, focusing on the adiabatic regime $\kappa_b \gg g_2 |\alpha|$ (purple curves in Fig.~\ref{fig:liouvillian_gap_dissipative}).
The overlayed trajectory in Fig.~\ref{fig:single_mode_traj} confirms that the memory undergoes the same four stages of the single-mode case.
We also remark that the photon number in the buffer mode remains small and we observe no sizable coherent exchange between memory and  buffer (not shown).

\subsection{Nonadiabatic and linear two-mode cats}
\label{sec:nonadiabatic_linear_bit_flips}

We now investigate deviations from the single-mode physics in
Fig.~\ref{fig:liouvillian_gap_dissipative} (a) considering $\kappa_b \simeq g_2 |\alpha|$ and linear buffer and memory modes ($K_a = K_b = \chi = 0$), in the presence of both photon loss and dephasing.
This corresponds to the bit-flip error rates in Fig.~\ref{fig:liouvillian_gap_dissipative} (a).

\begin{figure}[t]
\centering
\includegraphics[width=0.48 \textwidth]{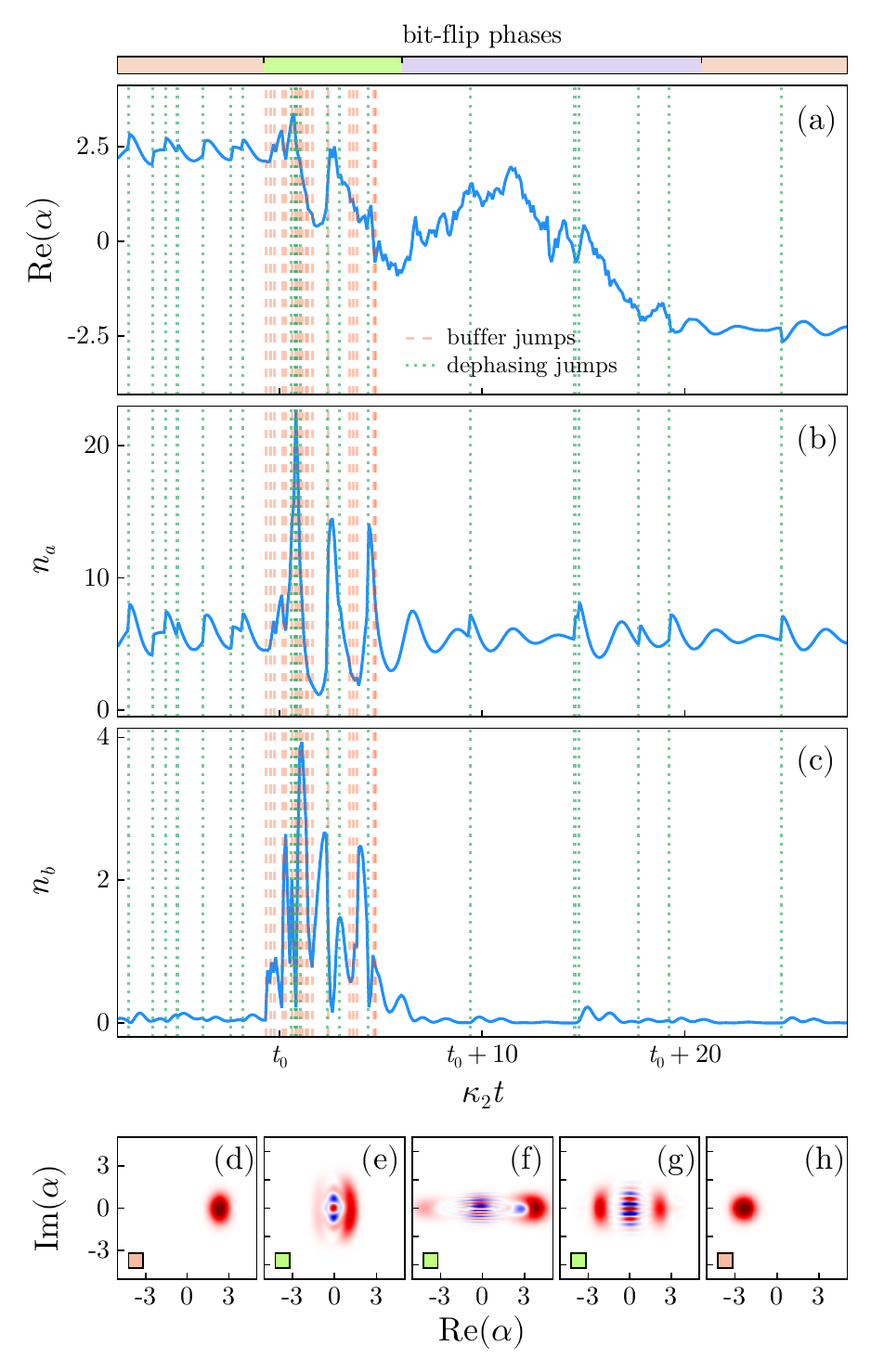}\vspace{0.02em}
\caption{Nonadiabatic two-mode bit-flip.
Panels (a,b) and the colorbar report the same quantities as in Fig.~\ref{fig:single_mode_traj}. 
Panel (c) show the photon number in the buffer $\langle\Psi(t)|\hat{b}^\dagger \hat{b}|\Psi(t) \rangle$.
(d-h) Wigner functions of the trajectory wave function $\ket{\Psi(t)}$ across the bit-flip event. Specifically, $W(\alpha,\,\alpha^*)$ for:
(d) the coherent phase preceding the bit-flip.
(e-g) the fluctuation phase, in correspondence of strong buffer emission and dephasing jumps.
(h) the coherent phase following the bit flip.
Parameters are $\kappa_b/2\pi = 1.44\,\mathrm{MHz}$ and $g_2/2\pi = 0.6\,\mathrm{MHz}$, while $K_b=\chi=0$.
$|\varepsilon_b|$ is chosen so that $\expval*{\hat{a}^\dagger \hat{a}}_{\rm ss} = 5.5$.
The other parameters are set as in Fig.~\ref{fig:single_mode_traj}.
}
\label{fig:two_mode_nonadiabatic_traj}
\end{figure}

We analyze the trajectory across a bit-flip event plotting the field quadrature, the photon number in the memory, and the photon number in the buffer in Figs.~\ref{fig:two_mode_nonadiabatic_traj} (a), (b), and (c), respectively.
The Wigner functions are shown in Figs.~\ref{fig:two_mode_nonadiabatic_traj} (d-h).
While the coherent phase shows similar properties to the adiabatic case [Fig.~\ref{fig:two_mode_nonadiabatic_traj} (d)], we observe several important differences.
First, coherent processes between the buffer and memory become visible and manifest themselves as oscillations of excitations between the two modes.
Second, the photon number in the memory no longer exhibits a purely vacuum phase but instead displays rapid oscillations between few and many photons.
Third, the buffer mode itself develops large fluctuations, and its photon number can become significantly larger than in the coherent phase.
Finally, we also notice that the Wigner functions in Figs.~\ref{fig:two_mode_nonadiabatic_traj} (e-g) across the bit flip are much more distorted than the ones reported in Figs.~\ref{fig:single_mode_traj} (d-f).

\begin{figure}[t]
\centering
\includegraphics[width=0.48 \textwidth]{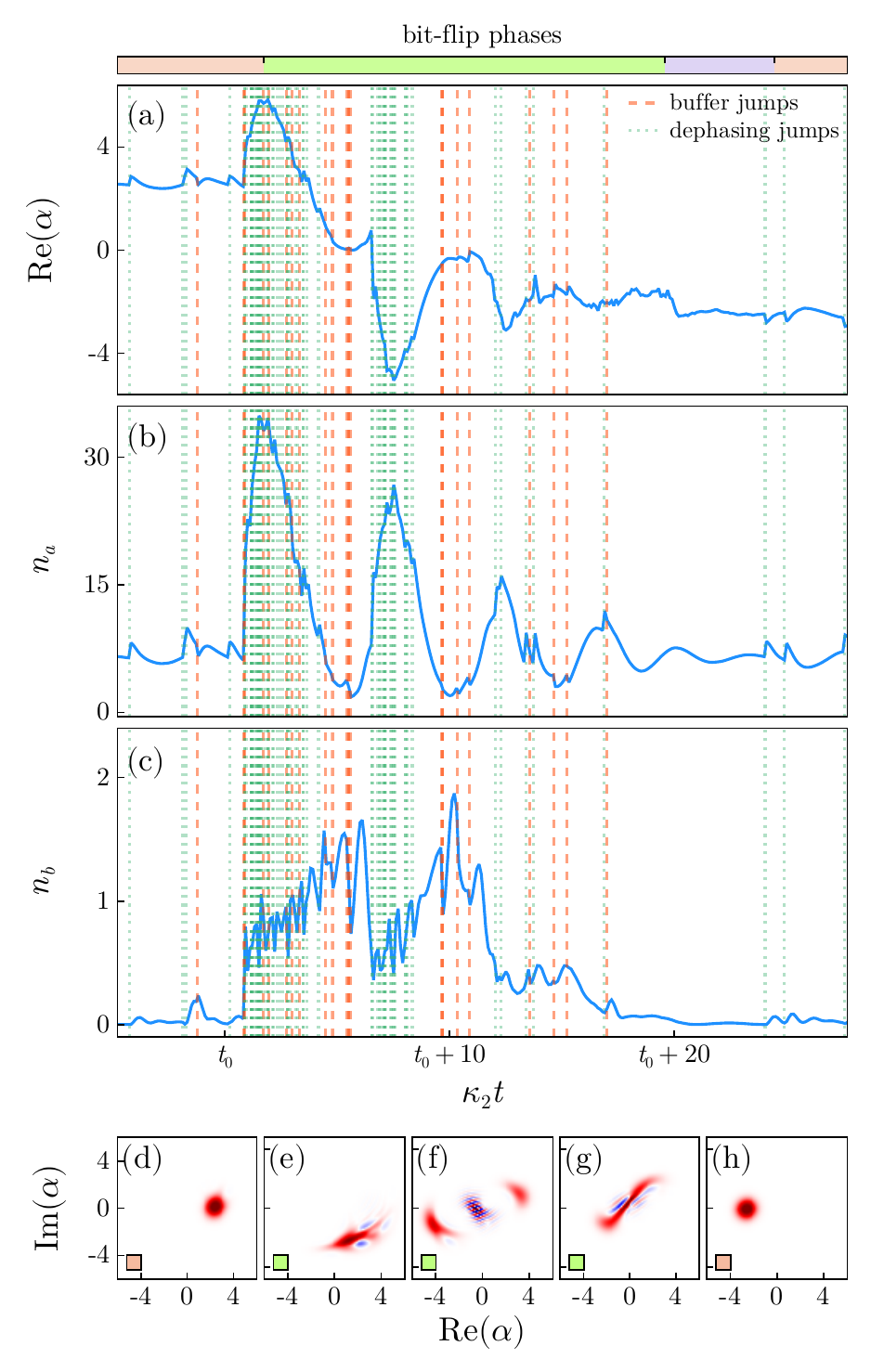}\vspace{0.02em}
\caption{Nonadiabatic two-mode bit-flip with nonzero Kerr and cross-Kerr nonlinearities in the buffer mode.
Same analysis over a single trajectory as in Fig.~\ref{fig:two_mode_nonadiabatic_traj} but for nonzero buffer nonlinearities: $K_b/2\pi=1\,\textrm{MHz}$ and $\chi/2\pi=0.5\,\textrm{MHz}$.
$|\varepsilon_b|$ is chosen so that $\expval*{\hat{a}^\dagger \hat{a}}_{\rm ss} = 6.5$.
The other parameters are fixed as in Fig.~\ref{fig:two_mode_nonadiabatic_traj}.
}
\label{fig:two_mode_nonlinear_traj}
\end{figure}

These observations indicate that the switching process is less structured than in the adiabatic regime and photon-number fluctuations are amplified through coherent exchanges between memory and buffer.
Rather than four clearly separated phases, the fluctuation and vacuum phases become intertwined, and the transition back to the cat manifold becomes less well defined.
Nonetheless, the dynamics is still regular, and the memory remain fixed along the $\textrm{Im}(\alpha)=0$ line in phase space.

\subsection{Nonadiabatic, nonlinear, two-mode cat}
\label{sec:nonadiabatic_linear_bit_flips_nonadiabatic_nonlinear}
\label{sec:nonadiabatic_linear_bit_flips_nonadiabatic}

In Fig.~\ref{fig:two_mode_nonlinear_traj} we consider $K_b, \, \chi \neq 0$ for the same the nonadiabatic regime studied in Sec.~\ref{sec:nonadiabatic_linear_bit_flips}.
This corresponds to bit-flip error rates that have saturated in Fig.~\ref{fig:liouvillian_gap_dissipative} (b).
As before, we plot the field and photon number in the memory, together with the photon number in the buffer, in Figs.~\ref{fig:two_mode_nonlinear_traj} (a), (b), and (c), respectively.
The Wigner functions across the bit flip are shown in Figs.~\ref{fig:two_mode_nonlinear_traj} (d-h).

Compared to Fig.~\ref{fig:two_mode_nonadiabatic_traj}, the switching process now displays qualitatively different features:
the dynamics exhibits strongly irregular oscillations and the topology of the jump change: the state in the memory becomes distorted,  explores a larger portion of phase space, and rotates in the $\textrm{Re}(\alpha)-\textrm{Im}(\alpha)$ plane as revealed by Figs.~\ref{fig:two_mode_nonlinear_traj} (d-h). 
The distinction between the different phases of the switching process is also further blurred.
It is the combined effect of cross-Kerr interactions and dephasing events (notice that many dephasing jumps occur during the first part of the dynamics) that strains the buffer and memory from the regular path discussed in the adiabatic but linear regime.
This mechanism of amplification of fluctuations ultimately leads to the saturation of the bit-flip rate identified in Sec.~\ref{sec:saturation}.

Importantly, far from bit-flip events the dynamics remains regular, and steady-state observables (i.e., averaged over many quantum trajectories) such as the photon number in both memory and buffer are almost indistinguishable from the value they take in the more regular regimes. 
In other words, the effects of cross-Kerr interactions and nonadiabaticity become manifest primarily during the switching process.
This observation highlights the importance of quantum-trajectory analysis.

\section{Semiclassical analysis, reflection symmetry, and phase locking}\label{sec:semiclassical_analysis}

\begin{figure}[t]
\centering
\includegraphics[width=0.49 \textwidth]{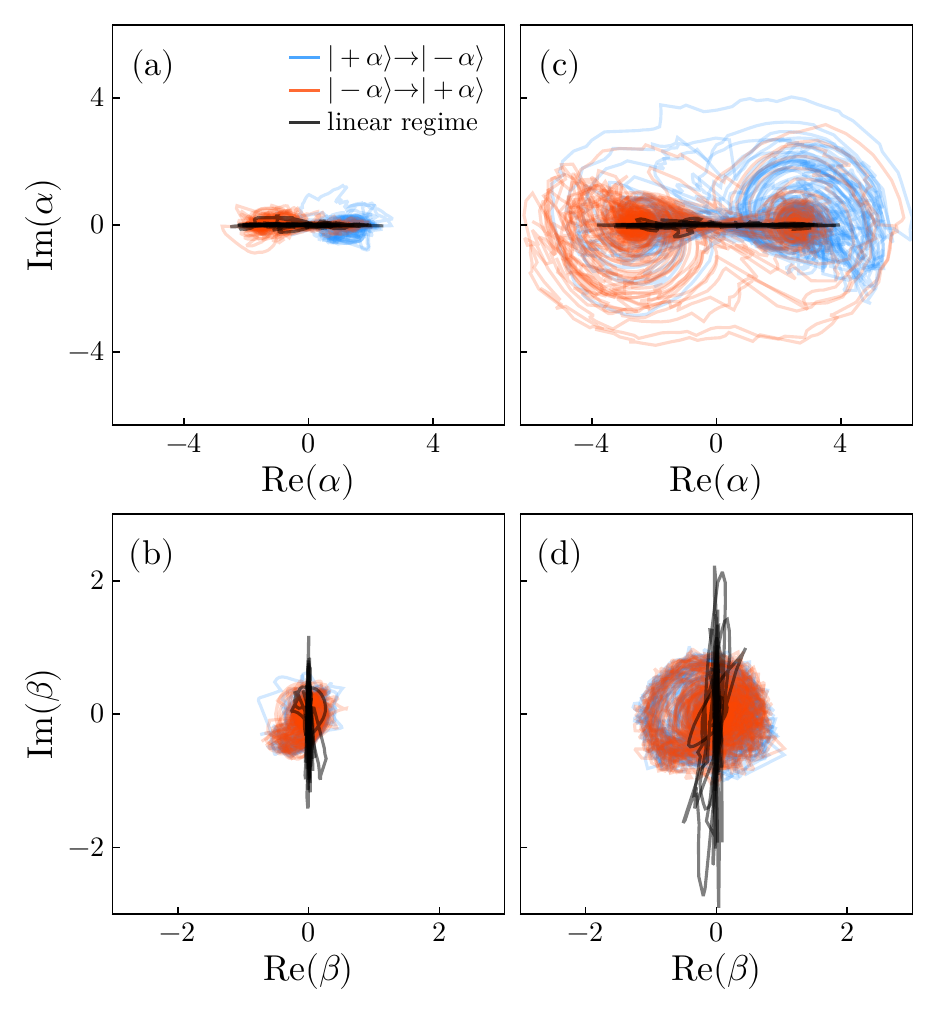}\vspace{0.01em}
\caption{Dynamics of $\alpha=\langle\Psi(t)|\hat{a}| \Psi(t)\rangle$ and $\beta=\langle \Psi(t)|\hat{b}|\Psi(t)\rangle$ for individual quantum trajectories $|\Psi(t) \rangle$ across the bit-flip for nonadiabatic cats.
Blue (red) curves indicate bit flips from $\ket{+\alpha}$ to $\ket{-\alpha}$ (from $\ket{-\alpha}$ to $\ket{+\alpha}$) for the nonlinear regime ($K_b/2\pi=1\,\textrm{MHz}$, $\chi/2\pi=0.5\,\textrm{MHz}$).
Black curves represent the bit flips in the linear regime ($K_b=\chi=0$).
In (a) and (b) we plot the dynamics of $\alpha$ and $\beta$ in the complex plane respectively for $\langle\hat{a}^\dagger\hat{a}\rangle=2.5$.
In (c) and (d) we plot the same quantities for $\langle\hat{a}^\dagger\hat{a}\rangle=6.5$.
The number of trajectories is fixed to 50 for the nonlinear case and to 5 for the linear case.
The other parameters are fixed as in Fig.~\ref{fig:two_mode_nonadiabatic_traj}.
}
\label{fig:bit_flip_dynamics}
\end{figure}

We now address the question of why, despite being nonadiabatic, the trajectories in Fig.~\ref{fig:two_mode_nonadiabatic_traj} can still display the good bit-flip scaling shown in Fig.~\ref{fig:liouvillian_gap_dissipative} (a), while it is only the combined role of nonadiabaticity and nonlinear interactions that breaks the cat's ability to suppress errors.

\subsection{Analysis of quantum trajectories}

In Fig.~\ref{fig:bit_flip_dynamics} we analyze $50$ trajectories showing bit flips and we plot $\alpha=\langle\Psi(t)|\hat{a}| \Psi(t)\rangle$ and $\beta=\langle \Psi(t)|\hat{b}|\Psi(t)\rangle$.
Black (colored) curves correspond to the linear (nonlinear) regime $\chi= K_b =0$ ($\chi, \, K_b \neq0$) in Fig.~\ref{fig:two_mode_nonadiabatic_traj} (Fig.~\ref{fig:two_mode_nonlinear_traj}).

At low photon numbers, where the adiabatic approximation remains valid both in the linear and nonlinear regime, the trajectory passes through the vacuum [see Fig.~\ref{fig:bit_flip_dynamics} (a)].
Correspondingly, the buffer responds fluctuating in the orthogonal direction [see Fig.~\ref{fig:bit_flip_dynamics} (c)].

At larger photon numbers in the memory, shown in Figs.~\ref{fig:bit_flip_dynamics} (c) and (d), adiabaticity breaks down. 
However, the linear model shows clear signature of a quasi phase locking between the two modes, and the motion of $\beta$ remains approximately confined to a one-dimensional manifold oriented at an angle $\pi/2$ with respect to the dynamics of $\alpha$.
On the contrary, the nonlinear system explores a much larger portion of phase space and trajectories can rotate around the coherent states [see Fig.~\ref{fig:bit_flip_dynamics} (c)].
These plots resemble the double-well pseudo-potential associated with cat states.
We also observe an asymmetry in the switching directions (from $\ket{+\alpha}$ to $\ket{-\alpha}$ and vice versa), indicating the breaking of time-reversal symmetry at the level of individual quantum trajectories \cite{carde_nonperturbative_2026}.
This is truly a four-dimensional dynamics for memory and buffer, and we verified that no locking takes place between $\alpha$ and $\beta$.
These results indicate that phase locking is a resource for cats dissipatively stabilized by a buffer mode, and show how nonlinear cross-Kerr interactions can break such a beneficial condition. 

\subsection{Semiclassical analysis}
\label{sec:semiclassical_universality}

To provide an intuition on the effects of nonlinearities, we use the semiclassical analysis.
In particular, we focus on cross Kerr.
The equations of motion can be obtained by approximating the system state as a product of coherent states $\hat\rho(t)=\ketbra{\alpha}\otimes\ketbra{\beta}$, with complex amplitudes \begin{equation}
    \alpha=x_a+\rmi y_a,\qquad \beta=x_b+\rmi y_b.
\end{equation}
describing the memory and buffer
fields, respectively.
They evolve through
\begin{equation}\label{eqs:semiclassical_cartesian}
\begin{split}
    \dot x_a
    &=
    -\frac{\kappa_a}{2}x_a
    -\chi|\beta|^2 y_a
    +2g_2\left(y_b x_a-x_b y_a\right),
    \\[2mm]
    \dot y_a
    &=
    -\frac{\kappa_a}{2}y_a
    + \chi|\beta|^2 x_a
    -2g_2\left(x_b x_a+y_b y_a\right),
    \\[2mm]
    \dot x_b
    &=
    -\frac{\kappa_b}{2}x_b
    - \chi|\alpha|^2 y_b
    +2g_2 x_a y_a,
    \\[2mm]
    \dot y_b
    &=
    -\frac{\kappa_b}{2}y_b
    + \chi|\alpha|^2 x_b
    -g_2\left(x_a^2-y_a^2\right)
    + \vert \varepsilon_b \vert.
\end{split}
\end{equation}

Eqs.~\eqref{eqs:semiclassical_cartesian} define a four–dimensional dynamical system.
Nevertheless, two limiting cases lead to effectively regular 2D reduced dynamics.
The first corresponds to the usual adiabatic regime, where the buffer variables $(x_b,y_b)$ rapidly adjust to the
memory variables $(x_a,y_a)$. This is discussed in Appendix~\ref{sec:adiabatic_approximation} and, roughly speaking, requires $\kappa_b \gg \chi (\varepsilon_b/\kappa_b)^2, \, g_2 \sqrt{n_a} $.
The second regime corresponds to phase locking, and we discuss it here.

\subsubsection{Linear system}
\label{sec:linear_semiclassical}
If $\chi =0$, the equations of motion are invariant under the reflection symmetry 
\begin{equation}
    x_a \mapsto x_a, \quad y_a \mapsto -y_a, \quad x_b \mapsto -x_b, \quad y_b \mapsto y_b.
\end{equation}
This symmetry leaves invariant the reflection plane $y_a=0,~ x_b=0$, which we call the phase-locked manifold. 
Indeed,
a system is initialized in ${y_a(t=0)=x_b(t=0)=0}$ remains on that plane, i.e., ${y_a(t)=x_b(t)=0}$.
This locking is true \textit{independently of the adiabaticity of the buffer}, reducing
the 4D system in Eqs.~\eqref{eqs:semiclassical_cartesian} in a 2D  motion for the memory $x_a(t)$ and buffer $y_b(t)$ coordinates reading \footnote{Notice that this further simplifies in an equation for the radiuses and relative sign of $\alpha$ and $\beta$. Introducing $u=\ln |x|$ one gets
$$
\ddot{u}-\frac{\kappa_b}{2} \dot{u}+2 g_2^2 e^{2 u}+\left(2 g_2 \varepsilon_b-\frac{\kappa_a \kappa_b}{4}\right)=0.
$$
}
\begin{equation}
\begin{split}
    &\dot x_a=
    x_a\left(2g_2 y_b-\frac{\kappa_a}{2}\right),\\
    &\dot y_b=
    -\frac{\kappa_b}{2}y_b-g_2x_a^2+\vert \varepsilon_b \vert,
\end{split}
\end{equation}
whose stable points are
\begin{equation}\label{eq:fixedpoints}
x_{a,\mathrm{st}} = \pm \sqrt{\frac{\vert \varepsilon_b \vert}{g_2}
-
\frac{\kappa_a\kappa_b}{8g_2^2}}, \qquad y_{b,\mathrm{st}}=\frac{\kappa_a}{4g_2}.
\end{equation}

\subsubsection{Effects of nonlinearity}

Nonlinearities break this symmetry.
A linearization procedure for the transverse variables $z_\perp=(y_a,
x_b)^\top$ around the locked plane gives
\begin{equation}
    \dot z_\perp=
    A_\perp z_\perp+s_{\rm pert},
\end{equation}
where $A_\perp \equiv A_\perp(x_a,y_b)$ is the Jacobian matrix (for a detailed derivation, see Appendix \ref{App:Stability_semiclassical}). 
Considering nonzero cross-Kerr, we get
\begin{equation}
    s_{\rm pert}=
    \begin{pmatrix}
    \chi y_b^2 x_a\\
    -\chi x_a^2 y_b
    \end{pmatrix}.
\end{equation}

As discussed in Appendix \ref{App:Stability_semiclassical}, the effect of this perturbation remains small in the limit 
\begin{equation}
    |\chi|\left(\frac{2\vert \epsilon_b\vert}{\kappa_b}\right)^2
    \ll
    \frac{\kappa_b}{4}, \qquad
    |\chi|  n_a
    \ll
    \frac{\kappa_b}{4},
\end{equation}
Namely, even outside of the adiabatic regime, the dynamics can be locked by $\kappa_b$, provided that the distortions induced by the cross-Kerr remain marginal with respect to it. 

\subsection{Symmetries and reminiscence of the phase-locked dynamics in the quantum system} 

Consider the ideal Liouvillian $\mathcal L_0$ in the absence of dephasing as well as Kerr and cross-Kerr nonlinearities.
Here, as in the previous section, we consider a real $\varepsilon_d<0$, so that the stabilized coherent states lie on the x-axis and typically satisfy $\alpha^2=-\varepsilon_d/g_2>0$. 
In analogy with the classical case, we introduce the quadratures
\begin{equation}
\begin{split}
    \hat{X}_a&=\hat{a}+\hat{a}^\dagger,
    \quad
    \hat{Y}_a=\rmi(\hat{a}^\dagger-\hat{a}),\\
    \quad
    \hat{X}_b&=\hat{b}+\hat{b}^\dagger,
    \quad
    \hat{Y}_b=\rmi(\hat{b}^\dagger-\hat{b}),
\end{split}
\end{equation}
and find that the antiunitary reflection 
\begin{equation}
\begin{split}
    \hat{X}_a\mapsto \hat{X}_a,
    \quad
    \hat{Y}_a\mapsto -\hat{Y}_a,
    \quad
    \hat{X}_b\mapsto -\hat{X}_b,
    \quad
    \hat{Y}_b\mapsto \hat{Y}_b,
\end{split}
\end{equation}
leaves the equation of motion invariant.

More formally, we can introduce the antiunitary reflection symmetry superoperator
\begin{equation}
    \mathcal T\hat{\rho}=\hat{\Theta}\hat{\rho}\hat{\Theta}^{-1}, \qquad \hat{\Theta}=\hat{\Pi}_b K
\end{equation}
where \(K\) denotes complex conjugation in the
Fock basis, and $\hat{\Pi}_b=(-\mathds{1})^{\hat{b}^\dagger \hat{b}}$ is the parity operator for the $b$ mode. One has
\begin{equation}
    \mathcal L_0\mathcal T= \mathcal T \mathcal L_0 \quad \Rightarrow \quad {[\mathcal L_0,\mathcal T]=0}.
\end{equation}
Since $\mathcal T$ has eigenvalues $\pm 1$, the Liouville space then decomposes into \(\mathcal T\)-even
and \(\mathcal T\)-odd sectors, defined by:
\begin{equation}
    \mathcal R_+
    =
    \{\rho:\mathcal T\rho=\rho\},
    \qquad
    \mathcal R_-
    =
    \{\rho:\mathcal T\rho=-\rho\}.
\end{equation}
Since transverse observables such as $\hat{Y}_a$ and $\hat{X}_b$ probe the $\mathcal T$-odd sector, any $\rho \in \mathcal R_+$ respects 
\begin{equation}
    \langle \hat{Y}_a\rangle=0,
    \qquad
    \langle \hat{X}_b\rangle=0 .
\end{equation}
However, even transverse fluctuations $\langle \hat{Y}_a^2\rangle$ and $\langle \hat{X}_b^2\rangle$, which are finite, can be modified.
Thus, the classical invariant manifold survives quantum mechanically not as a
zero-width constraint, but as a reflection-symmetric quantum tube around the
plane \(Y_a=X_b=0\).
In other words, the weak symmetry of the Liouvillian \textit{does not confine} the system to a sharp manifold in phase space, but only its expectations values. 

Given the Liouvillian symmetry, trajectories satisfy $\langle \hat{Y}_a\rangle=\langle \hat{X}_b\rangle=0 $ \textit{in average}.
However, for the unraveling considered here and for $\kappa_a =0$, both the non-Hermitian Hamiltonian and the $\hat{b}$ jump operator respect the same anti-unitary reflection symmetry, and so $\langle \hat{Y}_a\rangle=\langle \hat{X}_b\rangle=0 $ even along single quantum trajectories and during a bit-flip.
The term that is not invariant is the homodyne-like dynamics stemming from $\kappa_a$ (see Appendix \ref{sec:app_hom_traj}).
Nonetheless, as we are considering very small values of $\kappa_a$, bit-flip trajectories remain \textit{quasi phase locked} and satisfy $\langle \hat{Y}_a\rangle=\langle \hat{X}_b\rangle\simeq 0 $, as can be seen in Figs.~\ref{fig:two_mode_nonadiabatic_traj} and \ref{fig:bit_flip_dynamics}. 
The marginal fluctuations in Figs.~\ref{fig:two_mode_nonadiabatic_traj} and \ref{fig:bit_flip_dynamics}, captured by quantum trajectories and not by the semiclassical analysis, can be interpreted as the effect of quantum noise induced by dissipation \cite{wiseman_quantum_2009}.

In the presence of dephasing or Kerr nonlinearities, the system does not have this anti-unitary symmetry. The switching trajectories then bend away from $\langle \hat{Y}_a\rangle=\langle \hat{X}_b\rangle=0$ as it results from Figs.~\ref{fig:two_mode_nonlinear_traj} and \ref{fig:bit_flip_dynamics}.

\section{Signatures of chaotic behavior}\label{sec:dissipative_quantum_chaos}

\begin{figure*}[t]
\centering
\includegraphics[width=1 \textwidth]{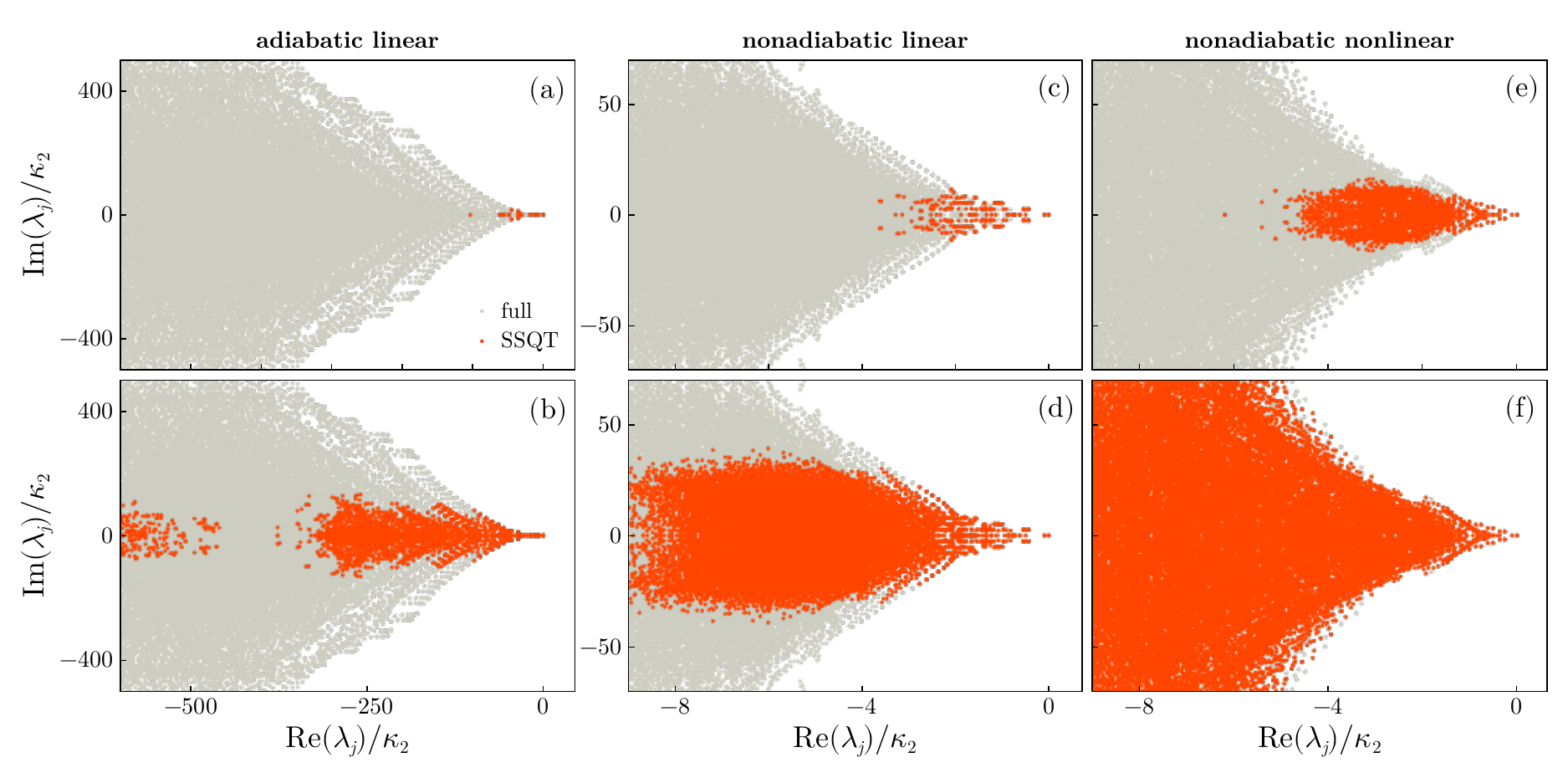}\vspace{0.02em}
\caption{Liouvillian eigenvalues selected by the SSQT criterion (a), (c), and (e), away from or (b), (d), and (f) during the bit-flip event.
The ivory-colored points indicate the full Liouvillian spectrum.
The red points indicate the Liouvillian eigenvalues.
(a-b) Adiabatic and linear two-mode cat.
The effective two-photon drive amplitude is fixed to $\varepsilon_2/\kappa_2=4$ and $g_2/2\pi=5\,\textrm{MHz}$.
(c-d) Nonadiabatic and linear two-mode cat for $\varepsilon_2/\kappa_2=4.5$ and $g_2/2\pi=0.6\,\textrm{MHz}$.
(c-d) Nonadiabatic and nonlinear two-mode cat for $\varepsilon_2/\kappa_2=5.5$ and $g_2/2\pi=0.6\,\textrm{MHz}$.
Furthermore, $K_a/2\pi=1\,\textrm{MHz}$ and $\chi/2\pi=0.5\,\textrm{MHz}$.
The Hilbert space cutoffs for the Liouvillian diagonalization are set to $N_a=24$ for the $a$ mode and $N_b=10$ for the $b$ mode.
}
\label{fig:ssqt}
\end{figure*}

The analysis in Secs.~\ref{sec:bit_flips} and \ref{sec:semiclassical_analysis} points to the fact that, when saturation occurs, the topology of bit-flip events changes.
We now connect these results to the emergence of dissipative chaos.

Chaos in cat qubits has been investigated mainly within the Kerr paradigm \cite{grimm_stabilization_2020} and a Hamiltonian framework \cite{haake_quantum_2018}.
Early works~\cite{chavez-carlos_spectral_2023, chavez-carlos_driving_2025, martinez_fundamental_2026}, showed that chaotic dynamics can steer the Kerr cat out of the logical manifold with chaos-assisted tunneling between the logical codewords.
The study of dissipative cat qubits and error correction, however, intrinsically requires an open quantum system approach~\cite{mirrahimi_dynamically_2014,putterman_hardware-efficient_2025}.

Various theories and tools~\cite{akemann_universal_2019, hamazaki_universality_2020, sa_complex_2020, dahan_classical_2022, prasad_dissipative_2022, kawabata_symmetry_2023, villasenor_breakdown_2024, ferrari_dissipative_2025, peyruchat_landauzener_2025} have been developed to study chaos in open quantum systems.
However, they are mainly based on the analysis of correlations in the Liouvillian spectrum~\cite{akemann_universal_2019, sa_complex_2020}; for the problem under consideration, where  $\kappa_a \ll g_2, \kappa_b$, we argue that these tools cannot be straightforwardly applied.
Indeed, the perturbative nature of $\kappa_a$ contaminates the spectral statistics by introducing nearly independent eigenvalues, making it challenging to identify the intrinsic spectral correlations of the system~\footnote{As detailed in, e.g., Ref.~\cite{sa_complex_2020}, spectral methods require separating the dynamics within symmetry sectors.
Otherwise, non-interacting eigenvalues would mix, resulting in spurious integrability.
Notice that, if $\kappa_a = 0$, the
system exhibits a \textit{strong} symmetry (i.e., conservation laws in the sense of Ref.~\cite{albert_symmetries_2014}), and the symmetry group of the Liouvillian is larger than the one with $\kappa_a \ne 0$.}. 
At the same time, $\kappa_a$ cannot be neglected, as it plays a crucial role in determining the occurrence of bit flips and their scaling. 
Following previous work on weakly dissipative systems~\cite{carlo_dissipative_2005}, we therefore adopt an alternative approach.
First, we want to characterize the typical timescales involved in a bit flip \cite{ferrari_dissipative_2025}.
Then, we assess the stochastic wave function delocalization in the Liouvillian spectrum~\cite{richter2025localizationdelocalizationquantumtrajectories}.
And finally, we quantify how entropic the Liouvillian eigenstates relevant for the bit-flip dynamics are.

\subsection{Activated eigenvalues}

To determine the complexity and time scales of a bit flip within a Liouvillian description, we cannot check arbitrary eigenvalues of $\LL$, but we rather have to select the portion of the spectrum which is needed to describe such an event. 
A procedure to fulfill this task is the spectral statistics of quantum trajectories (SSQT), formalized in Ref.~\cite{ferrari_dissipative_2025}.
The SSQT determines the \textit{activated} eigenvalues and eigenoperators of $\LL$, i.e., those required to describe the dynamics of single quantum trajectories showing the desired effect.
In Fig.~\ref{fig:ssqt} we plot the activated eigenvalues either in the coherent phase or during a bit flip, and compare them to the entire Liouvillian spectrum \footnote{The full diagonalization has been performed upon a rotation of the $a$ mode. Specifically, we consider the effective Hamiltonian $\hat{H}_{\rm eff} = \varepsilon_2\left(\hat{a}^2e^{-\rmi\theta} + \hat{a}^{\dagger 2}e^{+\rmi\theta}\right)/2 + \hat{a}^{\dagger 2}\hat{a}^2/2$, where $\theta$ controls the phase of the parity. 
We diagonalize $\hat{H}_{\rm eff}$, $\hat{H}_{\rm eff} = \sum_jE_j\ket{\Psi_j}\bra{\Psi_j}$, and we construct the rectangular matrix $\hat{R}_a = (\ket{\Psi_0}\,...\,\ket{\Psi_M})$.
We then perform the transformation on the $\hat{a}$ mode $\hat{a}\to\hat{R}_a^\dagger\hat{a}\hat{R}_a$. 
In the numerical simulations, we considered $M=200$ eigenstates of $\hat{H}_{\rm eff}$.}.
The other parameters are set as in Fig.~\ref{fig:single_mode_traj} (see Appendix \ref{sec:app_dqc} for more details).

In Figs.~\ref{fig:ssqt} (a) and (b) we consider the adiabatic case.
Between bit flips, the system evolution involves only few low-lying eigenvalues, showing that the system is regular [see Fig.~\ref{fig:ssqt} (a)].
When bit flips occur, more eigenvalues are activated which,
however, are such that Im$(\lambda_j)$/Re$(\lambda_j) \lesssim 1$ [see Fig.~\ref{fig:ssqt} (b)].
These processes are incoherent emissions, dominated by rapid decay rather than by coherent fluctuations.

In Figs.~\ref{fig:ssqt} (c) and (d) we then study the nonadiabatic and linear case.
Again, between bit flips the quantum trajectories activates few low-lying eigenvalues [see Fig.~\ref{fig:ssqt} (c)] and the dynamics is regular despite the breakdown of adiabaticity.
What significantly changes are the activated eigenvalues across the bit flip [see Fig.~\ref{fig:ssqt} (d)].
Now, a larger portion of the eigenvalues are selected, and for many eigenvalues Im$(\lambda_j)$/Re$(\lambda_j) \gg 1$.
We conclude that the involved processes are mainly coherent and oscillating. 
Let us remark here that the latter feature is not visible in all the observables.
Indeed, when comparing the data for the quadrature $\hat{a}$ to that of the photon number in the memory in Fig.~\ref{fig:two_mode_nonadiabatic_traj}, the former shows a regular dynamics, while the latter shows large and irregular fluctuations.
Connecting this result to the semiclassical analysis, we see that the phase-locking condition translates into a regular motion for $
\hat{a}$ and $\hat{b}$, with $\hat{a}^\dagger \hat{a}$ and $\hat{b}^\dagger \hat{b}$ representing $|x_a|$ and $|y_b|$ not being constrained.

\begin{figure}[t]
\centering
\includegraphics[width=0.48 \textwidth]{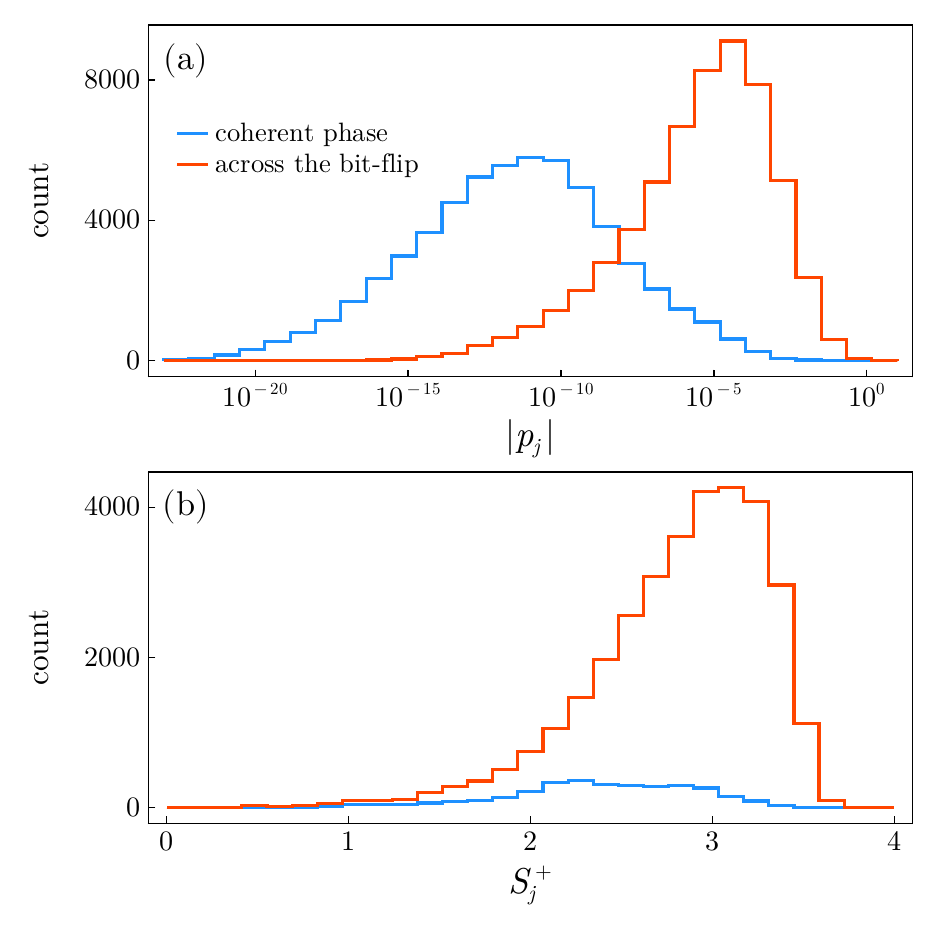}\vspace{0.02em}
\caption{Properties of the Liouvillian eigenstates in the nonadiabatic and nonlinear regime.
(a) Distribution of the quasi-probabilities $p_j$ (in log-scale) defined in Eq.~\eqref{eq:quasi_probabilities}.
The blue histograms refers the the dynamics between two bit-flips, while the red histograms refers to the dynamics across a bit-flip.
(b) Distribution of the entropy of the Liouvillian eigenstates $S_j^+ = -\operatorname{Tr}(\hat{\rho}_j^+\log\,\hat{\rho}_j^+)$ where $\hat{\rho}_j^+$ are density matrices defined from the right eigenvectors $\hat{\eta}_j$ as detailed in the Appendix \ref{sec:app_dqc}.
The eigenstates for which we plot $S_j^+$ are only the ones activated by the SSQT criterion.
Results are averaged over 10 quantum trajectories.
The other parameters are set as in Fig.~\ref{fig:single_mode_traj}.
}
\label{fig:eigenstates}
\end{figure}

Finally, in Figs.~\ref{fig:ssqt} (e) and (f) we study the nonadiabatic and nonlinear evolution.
In this case, between bit flips, single quantum trajectories activate more low-lying eigenvalues, but these are all compact around $\lambda_0 = 0$ and show little fluctuations [see Fig.~\ref{fig:ssqt} (e)].
Across the bit flip, instead, \textit{almost all} the Liouvillian eigenvalues are relevant, making it extremely irregular, and showing that no quantity remain predictable or well defined.

Such analysis indicates a change in the nature of bit-flip dynamics from integrable rare events to chaotic transient bursts interrupting an otherwise regular dynamics in the coherent phase. 

\subsection{Properties of the eigenstates}

To quantify the spreading of the wave function in the Liouvillian spectrum, we consider the quasi-probabilities $p_j$ introduced in Ref.~\cite{richter2025localizationdelocalizationquantumtrajectories}.
Given the quantum trajectory $\hat{\rho}(t) = \ketbra{\Psi(t)}$, we have the decompositions
\begin{equation}
    \hat{\rho}(t) = \sum_j c_j(t)\,\hat{\eta}_j=\sum_j d_j(t)\,\hat{\sigma}_j^\dagger,
\end{equation}
with $c_j(t) = \operatorname{Tr}[\hat{\sigma}_j^\dagger\hat{\rho}(t)]$ and $d_j(t) = \operatorname{Tr}[\hat{\eta}_j\hat{\rho}(t)]$.
The quasi-probabilities $p_j$ are defined as~\footnote{While the SSQT criterion is mainly based on the right eigenvectors of the Liouvillian, with the normalization condition $\operatorname{Tr}(\hat{\eta}_j^\dagger\hat{\eta}_j)=1$, the quantities $p_j$ have the advantage of being independent of any normalization of the Liouvillian eigenstates.}
\begin{equation}\label{eq:quasi_probabilities}
    p_j = c_j d_j.
\end{equation}
In Fig.~\ref{fig:eigenstates} (b) we plot the histogram of the quasi-probabilities $p_j$, in the coherent phase (blue curve) and across the bit-flip (red curve).
In the coherent phase, the distribution of $|p_j|$ is peaked around $10^{-12}$, and the tail of the distribution is suppressed slightly above $10^{-5}$.
This shows that $\ket{\Psi(t)}$ does not spread over many Liouvillian eigenvalues, in agreement with Fig.~\ref{fig:ssqt} (e).
Across bit flips the distribution of $|p_j|$ is peaked around $10^{-4}$, eight orders of magnitude larger than the previous case.
Furthermore, a significant fraction of quasi-probabilities is close to $10^{-2}-10^{-1}$, thus indicating a high degree of delocalization of the wave function in the Liouvillian spectrum, in line with Fig.~\ref{fig:ssqt} (f).
This diagnostic confirms the interpretation obtained above from the SSQT criterion.

Finally, we compute the von Neumann entropy of the activated Liouvillian eigenstates.
The reason for this choice is that, in closed quantum systems, the entanglement entropy of the system eigenstates, quantifying the mixedness of subsystems, is an established diagnostic of unitary thermalization and thus chaos~\cite{dalessio_quantum_2016}. 
In an open quantum system, where we traced out the environment, the mixedness of the Liouvillian eigenstates should then be a reliable diagnostic of chaotic-like behavior.

The entropy of the Liouvillian eigenstates $\hat{\eta}_j$ is ill defined, since they are traceless.
However, as we recall in Appendix \ref{sec:app_dqc}, each eigenstate can be decomposed as, e.g., $(\hat{\eta}_j + \hat{\eta}_j^\dagger)/2=\hat{\rho}_j^{+} - \hat{\rho}_j^{-}$ where $\hat{\rho}_j^{\pm}$ are positive operators with strictly nonzero traces of order unity. 
Upon normalization, they define proper density matrices, whose von Neumann entropy is
\begin{equation}\label{eq:eigenstate_entropy}
    S_j^{\pm} = -\operatorname{Tr}\!\left(\hat{\rho}_j^{\pm}\log\hat{\rho}_j^{\pm}\right).
\end{equation}

In Fig.~\ref{fig:eigenstates} (a) we plot the histogram of the entropies given by Eq.~\eqref{eq:eigenstate_entropy} associated with the activated eigenstates.
The blue curve describes the coherent phase, while the red one the bit-flip.
While in the coherent phase a small number of moderately entropic states are involved, across the bit flip thousands of more entropic states are now required to describe the dynamics.

Therefore, a large number of activated, \textit{entropic} eigenstates not only signals the delocalization of $\ket{\Psi(t)}$ across the spectrum, but also the structural complexity of the contributing eigenmodes themselves, supporting the interpretation of bit-flip as chaotic bursts in the saturation regime.

\section{Comparison with experiments}\label{sec:experiment}

We compare our theory with the data of Ref.~\cite{reglade_quantum_2024}.
The physical parameters are: $g_2/2\pi = 0.763\,\textrm{MHz}$, $\kappa_b/2\pi = 2.6\,\textrm{MHz}$ (yielding the ratio $\kappa_b/g_2\simeq3.4$, in between the barely- and almost-adiabatic regimes identified in Sec.~\ref{sec:saturation}), and $\kappa_a/2\pi=9.3\,\textrm{kHz}$.
Moreover, Ref.~\cite{reglade_quantum_2024} reports a memory thermal occupation of  $n_{\rm th}=0.1$.
To include this effect, we thus modify the single-photon loss rate as $\sqrt{\kappa_a}\to\sqrt{\kappa_a (1 + n_{\rm th})}$, and include the dissipator $\kappa_an_{\rm th}\mathcal{D}[\hat{a}^\dagger]$ to both Eqs.~\eqref{eqs:two_body} and \eqref{eq:single_body}.

Ref.~\cite{reglade_quantum_2024} does not provide explicit values for $\kappa_\phi$ and the nonlinearities $K_a$, $K_b$ and $\chi$.
To estimate the dephasing rate, we use the phase-flip data in Ref.~\cite{reglade_quantum_2024}: the slope of $\Gamma_{\rm pf}$ as a function of $\langle\hat{a}^\dagger\hat{a}\rangle$ exceeds the theoretical prediction $2\kappa_a$ by an amount compatible with $\kappa_\phi/2\pi\simeq1\,\textrm{kHz}$.
To estimate the remaining nonlinearities, we consider the full circuit Hamiltonian including external fluxes and zero-point fluctuations (see the Supplementary Information of Ref.~\cite{reglade_quantum_2024} and Ref.~\cite{berdou_one_2023} as well).
We get $K_a/2\pi\simeq6\,\textrm{Hz}\simeq0$, $K_b/2\pi\simeq3.3\,\textrm{MHz}$ and $\chi/2\pi\simeq0.28\,\textrm{MHz}$.
These estimates neglect pump-induced corrections as well as higher-order terms contributions and provide an indication about the order of magnitude of the above quantities.
Given the central role played by cross-Kerr, in the simulations, we consider multiple values of $\chi$ in the range $[0-0.3]\,\textrm{MHz}$.

\begin{figure}[t]
\centering
\includegraphics[width=0.48 \textwidth]{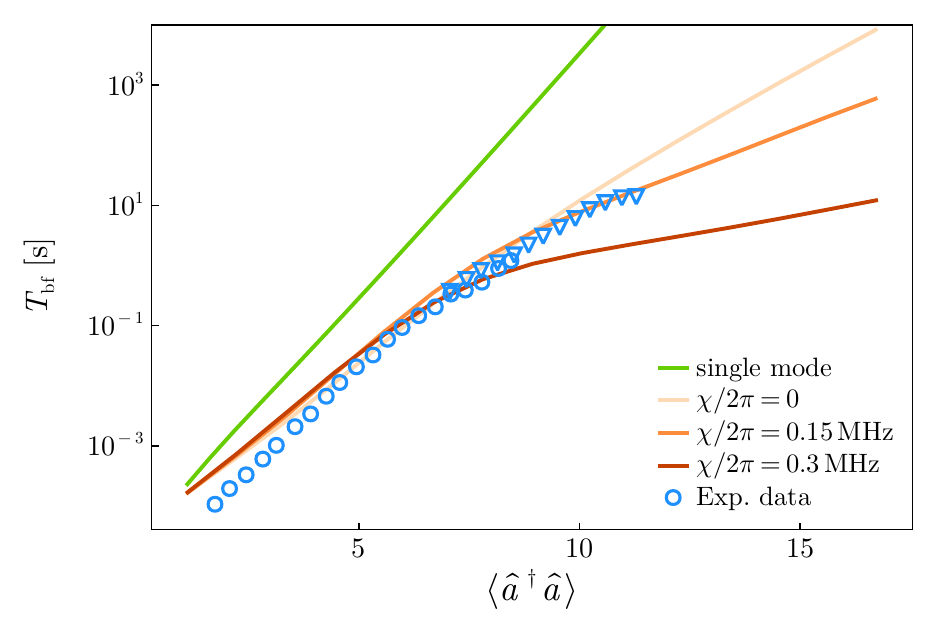}\vspace{0.02em}
\caption{Comparison with experimental data taken from Ref.~\cite{reglade_quantum_2024}.
The blue markers represent the measured bit-flip times in seconds.
The green lines indicates the simulation with the single-mode system given by Eq.~\eqref{eq:single_body}.
The orange lines correspond to the simulation of the full two-mode system given by Eqs.~\eqref{eqs:two_body} for three different values of $\chi$.
The physical parameters are
$g_2/2\pi = 0.763\,\textrm{MHz}$, $\kappa_b/2\pi = 2.6\,\textrm{MHz}$ (yielding $\kappa_2/2\pi=0.896\,\textrm{MHz}$), $\kappa_a/2\pi=9.3\,\textrm{kHz}$ and a thermal population in the memory of 10$\%$.
The remaining parameters are estimated to be $\kappa_\phi=1\,\textrm{kHz}$, $K_a=0$, $K_b/2\pi=3.3\,\textrm{MHz}$ and $\chi/2\pi = 0, 0.15, 0.3\,\textrm{MHz}$ (from light to dark orange). 
}
\label{fig:experimental_comparison}
\end{figure}

Results are presented in Fig.~\ref{fig:experimental_comparison}.
We plot the bit-flip time $T_{\rm bf} \propto 1/\Gamma_{\rm bf}$ as a function of the photon number $\langle\hat{a}^\dagger\hat{a}\rangle$.
The blue markers represent experimental data from Ref.~\cite{reglade_quantum_2024}.
The green line represents the single-mode approximation in Eq.~\eqref{eq:single_body}.
This model overestimates $T_{\rm bf}$ by more than four orders of magnitudes at large photon number, and shows no signatures of saturation, in agreement with Fig.~\ref{fig:liouvillian_gap_dissipative}.
The three orange lines, from light to dark, correspond to the two-mode model in Eqs.~\eqref{eqs:two_body} with cross-Kerr interactions $\chi=0$, $\chi/2\pi=0.15\,\textrm{MHz}$, and $\chi/2\pi=0.3\,\textrm{MHz}$.
All the two-mode simulations capture the correct order of magnitude of the measured bit-flip times: we confirm that a two-mode description is necessary to describe the cat qubit outside the adiabatic regime.
Finally, for $\chi/2\pi=0.15\,\textrm{MHz}$ we quantitatively retrieve the onset of saturation.

The comparison with experimental data establishes how the mechanisms described in this work predict the physics of dissipative cats and the onset of saturation, further highlighting the role of nonadiabaticity (all two-mode curves partially reproduce experimental data) and confirming the detrimental role of cross-Kerr interaction (larger cross-Kerr implies earlier onset of saturation).

\section{Conclusion}\label{sec:discussion}

We have presented a comprehensive framework to understand the dynamics of bit flips in dissipative cat qubits and the mechanisms responsible for the saturation of their error protection observed in experiments. 
By analyzing the full memory–buffer system, we demonstrated that bit-flip events are extended dynamical processes that intrinsically involve the coupled evolution of the two modes. 
As a consequence, the commonly used single-mode description fails to capture important features of the switching dynamics.

Our results reveal a hierarchy of dynamical regimes governing the behavior of dissipative cat qubits. In the adiabatic regime the system reproduces the regular switching dynamics predicted by the effective single-mode model. 
Remarkably, even when adiabaticity breaks down, the dynamics can remain regular due to the presence of a reflection symmetry of the Liouvillian, manifesting as a quasi phase-locking condition between the memory and buffer modes in quantum trajectories. 
This phase locking effectively constrains the switching trajectory and preserves the exponential suppression of bit-flip errors with increasing photon number.
When nonlinear interactions, dephasing, and nonadiabatic effects act simultaneously, there is no Liouvillian symmetry and the locking mechanism breaks down. 
In this regime fluctuations between the two modes are amplified, the switching trajectory explores a much larger region of phase space, and the suppression of bit flips saturates. 
We then analyze the dynamics of the system between (coherent phase) and during bit flips, to probe dissipative quantum chaos \cite{ferrari_dissipative_2025, richter2025localizationdelocalizationquantumtrajectories}.
We found that, once saturation occurs, bit flips become rare chaotic-like bursts interrupting regular dynamics.

We complete the treatment by comparing our theoretical predictions with the measured bit-flip times reported in Ref.~\cite{reglade_quantum_2024}, finding quantitative matching when the full two-mode structure of the device is considered, highlighting the role of nonadiabaticity and cross-Kerr interaction.

Overall, our results provide a unified framework not only for understanding the experimentally observed saturation of bit-flip times and clarify the dynamical mechanisms that limit the performance of dissipative cat qubits, but more in general to discuss chaos in bosonic codes, and the limitation it imposes on error correction protocols.
More pragmatically, these findings provide design guidelines for future cat-qubit architectures and open new directions for exploring and mitigating the interplay between quantum error correction, nonlinear dynamics, and dissipative quantum chaos in bosonic systems.

\begin{acknowledgments}
We acknowledge enlightening discussions with P. Campagne-Ibarcq, G. Campanaro, A. Melo, A. Mercurio and P. Winkel.
V. S. and F. F. acknowledge support by the Swiss National Science Foundation through Projects No. 200020\_215172, 200021-227992, and 20QU-1\_215928, and as part of NCCR SPIN (grant number 225153).
\end{acknowledgments}

\appendix

\section{Stochastic quantum trajectories}
\label{sec:Quantum_trajectories}
\label{sec:app_hom_traj}

The equation of motion of quantum trajectories can be derived assuming a perfect detector, i.e., assuming a ``click-no click'' dynamics, and conditioning the wave function to the presence or absence of each quantum jump \cite{dalibard_wave-function_1992, molmer_monte_1993} . 
Each quantum jump occurs in a time step $\rmd t$ with probability $\rmd p_\mu =  \langle\hat{L}_\mu^{\dagger}\hat{L}_\mu\rangle\,\rmd t$, and the wave function $\ket{\Psi(t)}$ evolves into
\begin{equation}\label{eqs:jump_operators}
    \ket{\Psi(t +\rmd t)}  \propto \hat{L}_\mu\ket{\Psi(t)}.
\end{equation}
No quantum jump occurs in the time $\rmd t$ with probability $1-\rmd \sum p_j$, and $\ket{\Psi(t)}$ evolves into
\begin{equation}\label{eqs:non_hermitian_hamiltonian}
    \ket{\Psi(t + \rmd t)} \propto (\mathds{1} - \rmi \, \rmd t\hat{H}_{\rm nh})\ket{\Psi(t)},
\end{equation}
where $\hat{H}_{\rm nh} = \hat{H} -\rmi  \sum_\mu \hat{L}_\mu^{\dagger}\hat{L}_\mu/2$ is the associated non-Hermitian Hamiltonian. 
After each time step, the state is renormalized according to $\ket{\Psi(t+\rmd t)} \mapsto \ket{\Psi(t+\rmd t)}/\langle\Psi(t+\rmd t)|\Psi(t+\rmd t)\rangle$. 
The overall process leads to a stochastic Schr\"odinger equation for the wave function $\ket{\Psi(t)}$.

The homodyne measurement of a specific jump operator $\hat{L}_\mu$ is the result of the same ideal detection scheme where, however, the incoherent, dissipated field is mixed with a constant coherent field of amplitude $\beta$.
While in many textbooks the limit $\beta \to \infty$ is taken, and the stochastic evolution results in a Wiener process for the system wave function, here we consider the effect of finite but large values of $\beta$, namely
\begin{equation}\label{eqs:beta-dyne}
    \hat{H}\to\hat{H} - \rmi(\hat{L}_\mu\beta^* - \beta\hat{L}_\mu^\dagger)/2,\qquad \hat{L}_\mu\to\hat{L}_\mu+\beta.
\end{equation}
For finite $\beta$, and with the same procedure detailed above for Monte Carlo trajectories, Eqs.~\eqref{eqs:beta-dyne} result in frequent quantum jumps, but whose effects on the  wave function are small, leading to a quantum state diffusion or quantum Brownian motion.
This $\beta$-dyne quantum trajectory \cite{Minganti_2018} has the advantage of being easy to simulate using advanced integration algorithms.

As discussed in Ref.~\cite{bartolo_homodyne_2017}, to observe bit flips it is necessary to preform a homodyne detection scheme of the output filed of the memory.
The mathematical reason for this can be understood using the single mode model.
The model we consider shows a weak  $\mathbb{Z}_2$ symmetry. Namely, $\hat{\Pi} \hat{\rho}(t) \hat{\Pi}  = \hat{\rho}(t)$ with $\hat{\Pi} = \exp(\rmi \pi \hat{a}^\dagger \hat{a})$.
Then, given a counting quantum trajectory initialized in an even cat, such that $\hat{\Pi} \ket{\Psi(0)}= (+1) \ket{\Psi(0)}$, at any time $t$ we have
\begin{equation}
\hat{\Pi} \ket{\Psi(t)} = (-1)^{J_a}\ket{\Psi(t)},
\end{equation}
where $J_a$ is the number of photon loss jumps.
The proof of this property is straightforward and can be adapted from \cite{Minganti_2018}.
It follows that, at all time, the system is in an eigenstate of the parity operator, and thus it will never explore the states $\ket{\pm \alpha}
$ and thus it will never show bit flips.
A homodyne (or $\beta$-dyne) trajectory, instead, does not obey this property, as $\left[\hat{H}_{\rm eff}, \hat{\Pi}\right] \neq 0$, thus making it possible to witness bit-flip events.

\section{Additional details about single and two-mode systems}

\subsection{Single-mode systems}\label{sec:appendix_single_mode}

\subsubsection{Bit-flips without dephasing}\label{sec:appendix_single_mode_nodeph}

\begin{figure}[t]
\centering
\includegraphics[width=0.48 \textwidth]{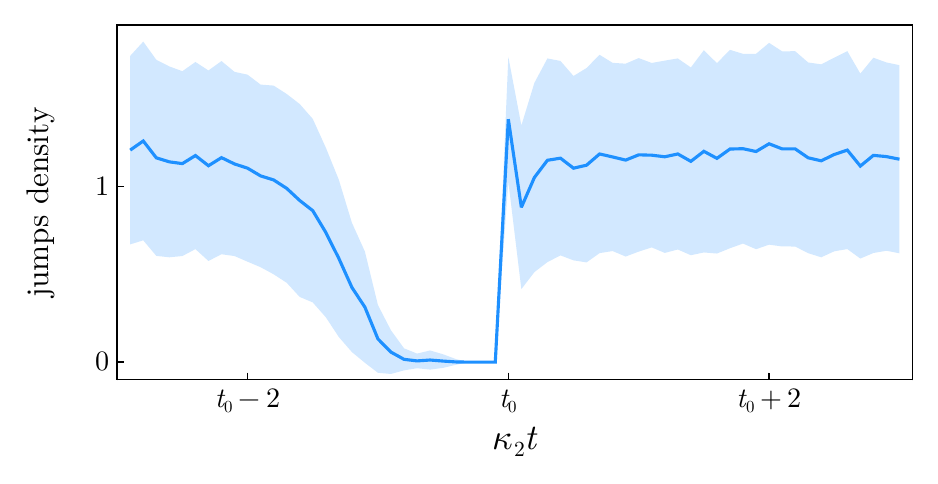}\vspace{0.01em}
\caption{As in Fig.~\ref{fig:jump_statistics} (a), average occurrence of two-photon quantum jump across bit-flip events in the single mode system described by Eq.~\eqref{eq:single_body}, but for $\kappa_\phi=0$ and $\kappa_a/2\pi=100\,\textrm{kHz}$.
Data are collected over $10^3$ trajectories exhibiting a bit-flip.
Other parameters are set as in Fig.~\ref{fig:single_mode_traj}.
}
\label{fig:jump_statistics_nodeph}
\end{figure}

We study the bit-flip mechanism for the single-mode model for $\kappa_\phi=0$. 
To compensate for the lower bit-flip rate, we increase the single-photon loss rate to $\kappa_a/\kappa_2=0.1$ and we slightly decrease the two-photon drive amplitude to have $
\langle\hat{a}^\dagger \hat{a}\rangle = 3.5$ ($\varepsilon_2/\kappa_2\simeq3.5$).
We keep all the other parameters as in Fig.~\ref{fig:jump_statistics}.
In Fig.~\ref{fig:jump_statistics_nodeph} we plot the two-photon jump density averaged over 1000 trajectories.
Similarly to Fig.~\ref{fig:jump_statistics}, we observe a suppression of two-photon jumps around the bit-flip event.

\begin{figure}[t]
\centering
\includegraphics[width=0.48 \textwidth]{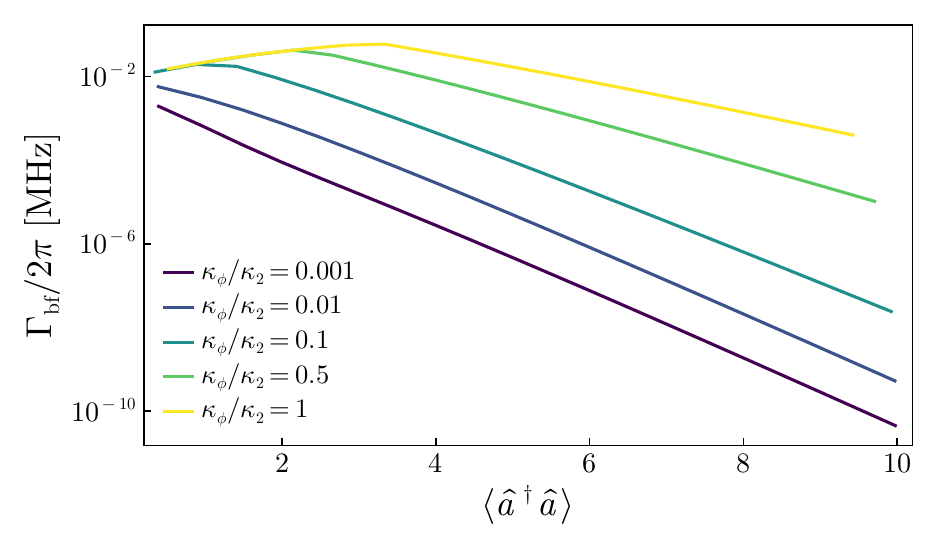}\vspace{0.01em}
\caption{Bit-flip error rate in the single mode system described by Eq.~\eqref{eq:single_body} with $K_a=0$ as a function of the photon number $\langle\hat{a}^\dagger\hat{a}\rangle$ for increasing values of $\kappa_\phi/\kappa_2$.
Other parameters are set as in Fig.~\ref{fig:single_mode_traj}.
}
\label{fig:liouvillian_gap_single_deph}
\end{figure}

\begin{figure}[t]
\centering
\includegraphics[width=0.48 \textwidth]{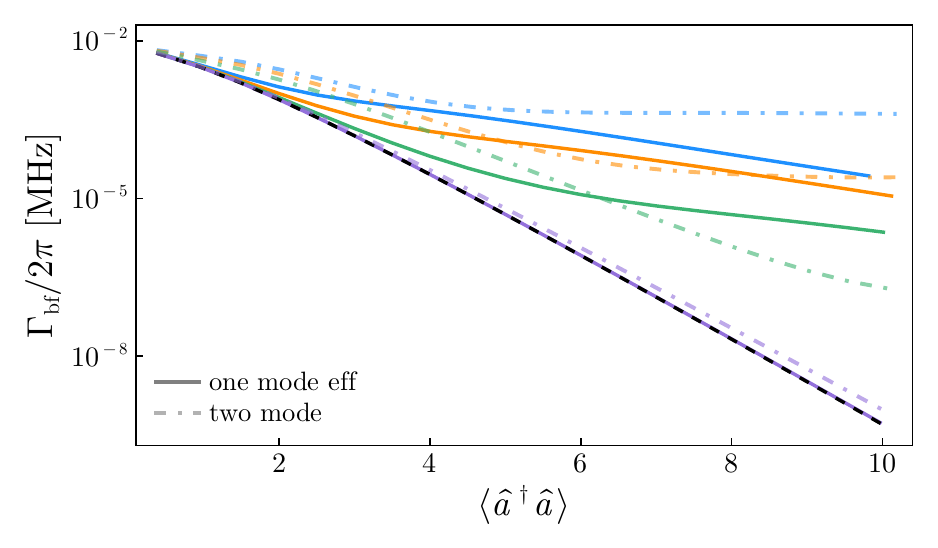}\vspace{0.01em}
\caption{
Bit-flip error rate $\Gamma_{\rm bf}$ in the single mode system derived from the effective operator formalism [according to Eqs.~\eqref{eqs:effective_operator}] with $K_a=K_b=0$ as a function of the photon number $\langle\hat{a}^\dagger\hat{a}\rangle$.
The memory single-photon and dephasing jump operators remain unchanged.
The dephasing rate is set to $\kappa_\phi/2\pi=10\,\textrm{kHz}$.
The black-dahsed line indicates the single-mode approximation of Eq.~\eqref{eq:single_body}.
Other parameters are set as in Fig.~\ref{fig:liouvillian_gap_dissipative}, and same color schemes for the curves.
}
\label{fig:liouvillian_gap_effective_operator}
\end{figure}

\begin{figure*}[t]
\centering
\includegraphics[width=0.8 \textwidth]{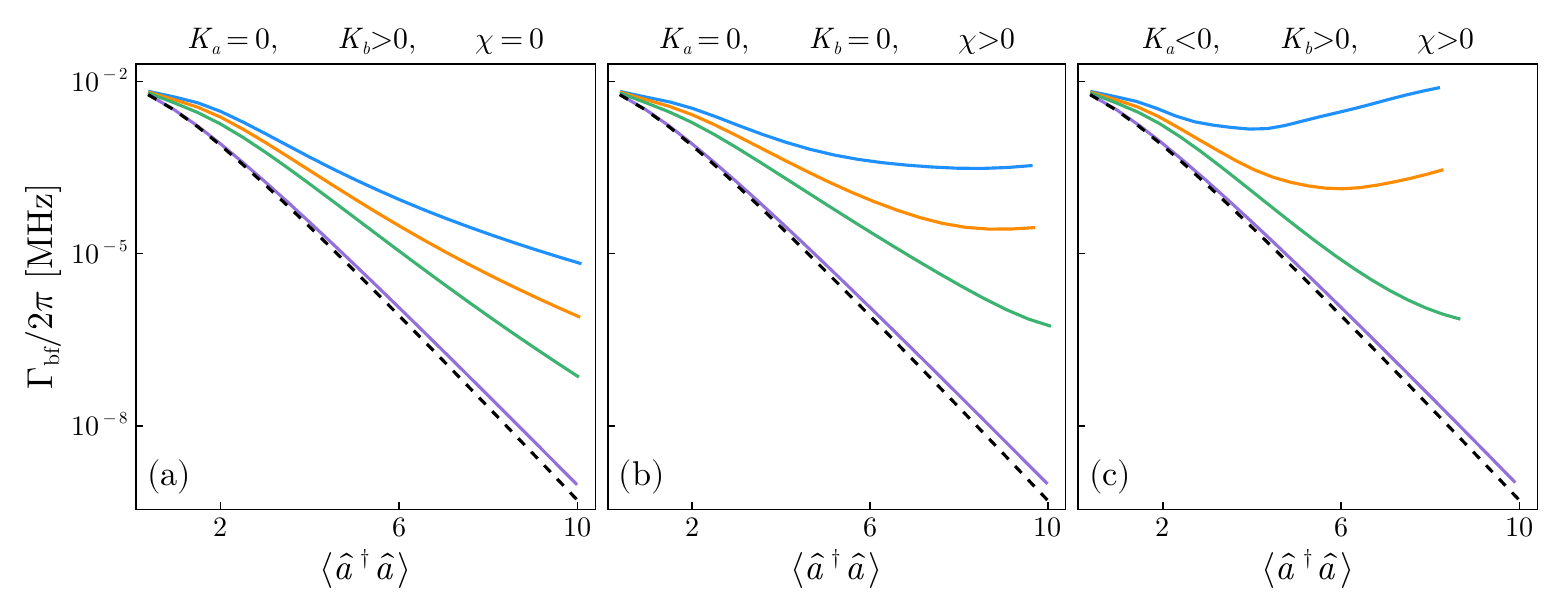}\vspace{0.01em}
\caption{
Bit-flip error rate $\Gamma_{\rm bf}$ as a function of the photon number in the memory for dissipative cat qubits with additional configurations of $(K_a,\,K_b, \,\chi)$ beyond those considered in Fig.~\ref{fig:liouvillian_gap_dissipative}.
In particular, we consider: (a) $K_b/2\pi = 1\,\textrm{MHz}$ and $\chi=0$ for $\theta=0$; (b) $K_b= 0$ and $\chi/2\pi=0.5\,\textrm{MHz}$ for $\theta=0$;
(c) $K_b/2\pi = 1\,\textrm{MHz}$, $\chi/2\pi=0.5\,\textrm{MHz}$ and $K_a<0$ for $\theta=\pi/32$.
Here the parameter $\theta$ is defined according to Eqs.~\eqref{Eq:nonlinearities}.
The dephasing rate is set to $\kappa_\phi/2\pi = 10\,\textrm{kHz}$.
Other parameters are set as in Fig.~\ref{fig:liouvillian_gap_dissipative}, and same color schemes for the curves.
}
\label{fig:liouvillian_gap_dissipative_culprit}
\end{figure*}

\subsubsection{Bit-flip error rate for increasing dephasing}\label{sec:appendix_single_mode_lergedeph}

We now show that the saturation effect unveiled in Sec.~\ref{sec:saturation} cannot be recovered within a simplified single-mode picture with an enhanced dephasing rate $\kappa_\phi$.
In Fig.~\ref{fig:liouvillian_gap_single_deph} we plot the bit-flip error rate $\Gamma_{\rm bf}$ as a function of the memory photon number $\langle\hat{a}^\dagger\hat{a}\rangle$ for the single-mode cat in Eq.~\eqref{eq:single_body} for increasing $\kappa_\phi = \kappa_2/1000$ to $\kappa_\phi=\kappa_2$, but for the same parameters as in Fig.~\ref{fig:single_mode_traj}.
The results indicate that $\Gamma_{\rm bf}$ always decays exponentially with the drive amplitude and never exhibits a saturation similar to that presented in Fig.~\ref{fig:liouvillian_gap_dissipative} (b).

\subsubsection{Bit-flip error rate within effective operator formalism}\label{sec:appendix_single_mode_effective}

We show that the bit-flip saturation cannot be captured through systematic perturbation theory based on the elimination of the $b$ mode.
Specifically, we use the effective operator formalism initially developed in Ref.~\cite{reiter_effective_2012} and applied to dissipative cat qubits in Ref.~\cite{putterman_preserving_2025}.
Following Refs.~\cite{reiter_effective_2012, putterman_preserving_2025}, the non-Hermitian Hamiltonian of the excited-state manifold is $\hat{H}_{\rm NH}=\hat{H}_{\rm e} - \rmi\hat{L}^\dagger\hat{L}/2$ with $\hat{H}_{\rm e}=\chi\hat{a}^\dagger\hat{a}$ and $\hat{L}=\sqrt{\kappa_b}$.
The effective Hamiltonian and jump operator in the ground-state manifold then read $\hat{H}_{\rm eff} = -\frac{1}{2}\hat{V}_-\left[\hat{H}_{\rm NH}^{-1}+(\hat{H}_{\rm NH}^{-1})^\dagger\right]\hat{V}_+$ and $\hat{L}_{\rm eff}=\hat{L}\hat{H}_{\rm NH}^{-1}\hat{V}_+$, being $\hat{V}_+ = g_2(\hat{a}^2-\alpha^2)=\hat{V}_-^\dagger$ and $\alpha=\varepsilon_2/\kappa_2$.
We have \cite{putterman_preserving_2025}
\begin{equation}
    \hat{H}_{\rm NH}^{-1} = \sum_{n=0}^\infty\frac{2\rmi}{\kappa_b+2\rmi\chi n}\ketbra{n},
\end{equation}
and thus:
\begin{align}\label{eqs:effective_operator}
    &\hat{H}_{\rm eff} = -\frac{g_2^2}{2}\left(\hat{a}^{\dagger2}-\alpha^2\right)\sum_{n=0}^\infty\frac{8\chi n}{\kappa_b^2+4\chi^2 n^2}\ketbra{n}\left(\hat{a}^2-\alpha^2\right),\nonumber\\
    &\hat{L}_{\rm eff} = \frac{2\rmi g_2}{\sqrt{\kappa_b}}\sum_{n=0}^\infty\frac{\kappa_b}{\kappa_b+2\rmi\chi n}\ketbra{n}\left(\hat{a}^2-\alpha^2\right).
\end{align}
Within this treatment, all the jump operators on the $a$ mode remain unchanged.

In Fig.~\ref{fig:liouvillian_gap_effective_operator} we plot the bit-flip error rate $\Gamma_{\rm bf}$ as a function of the memory photon number $\langle\hat{a}^\dagger\hat{a}\rangle$ .
All two-mode parameters are chosen to replicate Fig.~\ref{fig:liouvillian_gap_dissipative} (b) with $\kappa_\phi\ne0$.
We observe that saturation is not captured by the effective single-mode description.
This is in line with the quantum trajectory analysis, where we show that saturation occurs in regimes where bit flips become two-mode dynamical processes, in which memory and buffer coherently exchange excitations on timescales comparable to the switching itself.
This makes the use of static correction not predictive, regardless of the order of the expansion, confirming that the saturation of the bit-flip error rate is a genuine signature of the two-mode nonlinear dynamics identified in Secs.~\ref{sec:semiclassical_analysis} and \ref{sec:dissipative_quantum_chaos}.

\subsection{Two-mode systems}\label{sec:appendix_two_mode}

\subsubsection{Role of cross-Kerr and Kerr in the buffer}\label{sec:appendix_two_mode_role}

Figure~\ref{fig:liouvillian_gap_dissipative_culprit} further clarifies the role $\chi$ and $K_b$  in the scaling of $\Gamma_{\rm bf}$ with respect to Fig.~\ref{fig:liouvillian_gap_dissipative}.
Fig.~\ref{fig:liouvillian_gap_dissipative_culprit} (a) shows the scaling of $\Gamma_{\rm bf}$ with $\langle\hat{a}^\dagger\hat{a}\rangle$ when $\chi=0$ and $K_b/2\pi=1\,\textrm{MHz}$.
In this case no saturation is observed and the system qualitatively behaves as in Fig.~\ref{fig:liouvillian_gap_dissipative} (b) (completely linear buffer mode).
This result is in line with Ref.~\cite{gautier_combined_2022}, where the two-photon exchange Hamiltonian was introduced.
Fig.~\ref{fig:liouvillian_gap_dissipative_culprit} (b) shows the scaling of $\Gamma_{\rm bf}$ when $\chi/2\pi=0.5\,\textrm{MHz}$ and $K_b=0$.
Here, instead, saturation is recovered when adiabaticity breaks down.
We therefore claim that the fluctuations induced by the cross-Kerr interaction alone can cause this phenomenon.
This is consistent with the prediction of the semiclassical analysis discussed in Sec.~\ref{sec:semiclassical_analysis}.

\subsubsection{Nonzero memory Kerr with positive sign}\label{sec:appendix_kerr_memory_positive}

\begin{figure}[t]
\centering
\includegraphics[width=0.48 \textwidth]{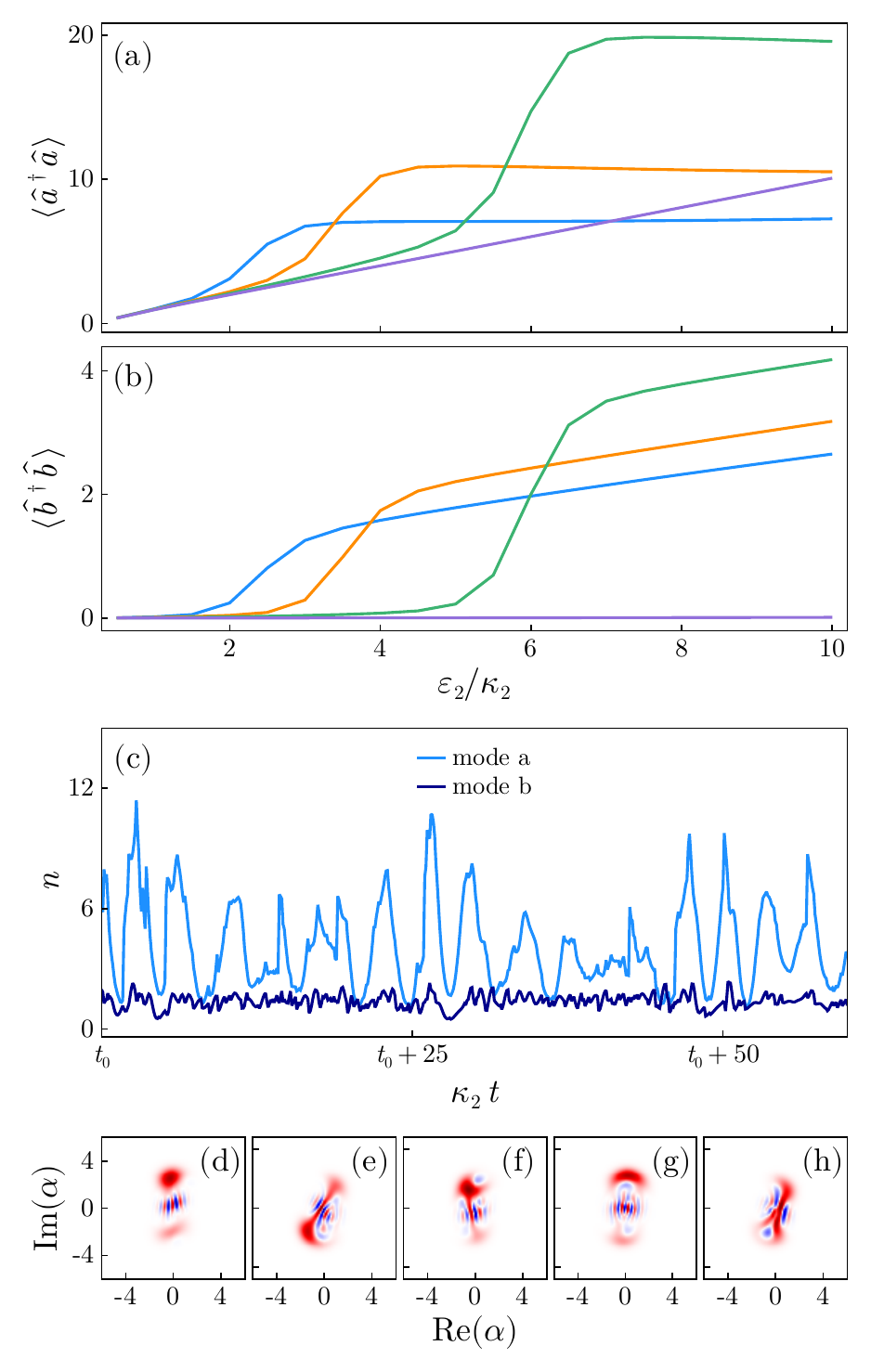}\vspace{0.01em}
\caption{Nonadiabatic two-mode quantum jumps with nonzero Kerr nonlinearity ($\theta=\pi/32$ according to Eq.~\eqref{Eq:nonlinearities}) in the buffer and in the memory, and with nonzero cross-Kerr. 
(a) Photon number in the memory $\langle\hat{a}^\dagger\hat{a}\rangle$ and 
(b) photon number in the buffer $\langle\hat{b}^\dagger\hat{b}\rangle$ as a function of the effective two-photon drive amplitude $\varepsilon_2/\kappa_2$ for the same values of $g_2$ used in Fig.~\ref{fig:liouvillian_gap_dissipative}.
(c) Dynamics of the photon number $n$ in the memory (light-blue curve) and in the buffer across a single quantum trajectory at $g_2/2\pi=0.6\,\textrm{MHz}$ and $\varepsilon_2/\kappa_2=5$.
(d-h) Wigner functions of the memory reduced density matrix, $\hat{\rho}_a(t) = \operatorname{Tr}_b\ketbra{\Psi(t)}$, along the dynamics of the quantum trajectory.
The other parameters are set as in Fig.~\ref{fig:two_mode_nonlinear_traj}.
}
\label{fig:two_mode_nonlinear_memory_traj}
\end{figure}

Here, we investigate more in details the mechanism leading to the degradation of the dissipative cat qubit highlighted in Figs.~\ref{fig:liouvillian_gap_dissipative} (c) when a memory Kerr with positive sign is considered.
In Figs.~\ref{fig:two_mode_nonlinear_memory_traj} (a) and (b) we plot the photon number in the memory $\langle\hat{a}^\dagger\hat{a}\rangle$ and in the buffer $\langle\hat{b}^\dagger\hat{b}\rangle$, respectively, as a function of the effective two-photon drive amplitude $\varepsilon_2/\kappa_2$ (i.e., for increasing drive $\varepsilon_b$).
Both resonators display, at a critical value of $\varepsilon_2/\kappa_2$, the onset of optical bistability \cite{drummond_quantum_1980, drummond_quantum_1981, bartolo_exact_2016}.
The desired steady-state manifold where the $a$ mode display a cat state and $b$ is in the vacuum shows the precursors of first-order dissipative phase transition \cite{minganti_spectral_2018} to a state with a higher population in both the modes.
Moreover, after the transition $\langle\hat{a}^\dagger\hat{a}\rangle$ becomes insensitive to $\varepsilon_2/\kappa_2$ while $\langle\hat{b}^\dagger\hat{b}\rangle$ linearly increases.
Notably, the critical point shifts towards smaller values of $\varepsilon_2/\kappa_2$ as adiabaticity breaks down.

The high-photon manifold does not host a cat qubit.
In Fig.~\ref{fig:two_mode_nonlinear_memory_traj} (c) we analyze the behavior of the photon number evaluated over a quantum trajectory of $n_a$ and $n_b$ in the memory and buffer, respectively. 
We observe a sequence of fast oscillations in $a$, indicating that no coherent states $\ket{\pm\alpha}$ is stabilized.
The analysis of the Wigner functions $W(\alpha,\,\alpha^*)$ of the reduced density matrix of the memory in Figs.~\ref{fig:two_mode_nonlinear_memory_traj} (d-h) confirms that the memory does not host a coherent state.

\subsubsection{Nonzero memory Kerr with negative sign}\label{sec:appendix_kerr_memory_negative}

Next, in Fig.~\ref{fig:liouvillian_gap_dissipative_culprit} (c) we investigate the role of a perturbative memory Kerr ($\theta=\pi/32$) with negative sign, $K_a<0$.
We assume $K_b/2\pi=1\,\textrm{MHz}$ and $\chi/2\pi=0.5\,\textrm{MHz}$ (nonlinear buffer with nonzero cross-Kerr).
As the memory is far detuned from the bistable branch, perturbative memory Kerr does not lead to the breakdown of the logical qubit.
Nonetheless, the saturation effect worsens with respect to the linear memory case analyzed in Fig.~\ref{fig:liouvillian_gap_dissipative} (b): instead of exhibiting a pleateau, the bit-flip error rate increases after a certain memory photon number $\langle\hat{a}^\dagger\hat{a}\rangle$, indicating that a larger cat degrades the autonomous error correction.
This is again consistent with the prediction of the semiclassical analysis discussed in Sec.~\ref{sec:semiclassical_analysis}.

\subsection{Hybrid cat qubits}\label{sec:appendix_hybrid_cats}

\begin{figure}[t]
\centering
    \begin{minipage}[c]{0.145\textwidth}
        \centering
        \includegraphics[width=\textwidth, trim={0.3cm 0 0.2cm 0.95cm}]{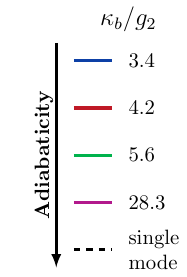} 
    \end{minipage}
    \begin{minipage}[c]{0.33\textwidth}
        \centering
        \includegraphics[width=\textwidth]{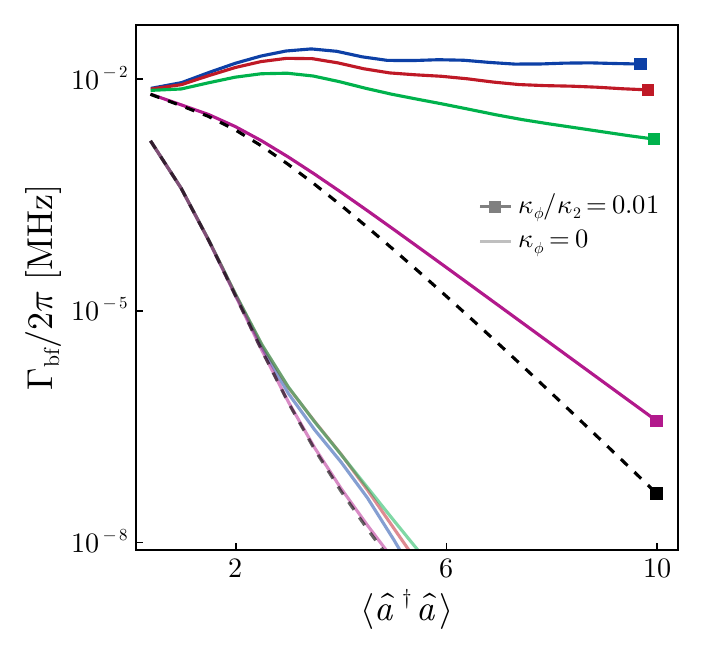}
    \end{minipage}
\caption{Bit-flip error rate $\Gamma_{\rm bf}$ as a function of the photon-number in the memory, but for a hybrid cat with $K_a = \kappa_2$ [$\theta=\pi/4$ according to Eq.~\eqref{Eq:nonlinearities}].  
Other parameters as in Fig.~\ref{fig:liouvillian_gap_dissipative} (a).
}
\label{fig:liouvillian_gap_hybrid}
\end{figure}

Finally, we briefly discuss the effect of non-adiabaticity on hybrid cat qubits, stabilized by an equal mixture of two-photon dissipation and Kerr nonlinearity. 
We will consider $K_b = \chi= 0 $ and the same values of $g_2/2\pi \in [0.6-5]\,$MHz as in Fig.~\ref{fig:liouvillian_gap_dissipative}, but for even larger span of $\kappa_b/2\pi \in [2.04-141]\,$MHz following Eq.~\eqref{Eq:nonlinearities}.

We report our findings in Fig.~\ref{fig:liouvillian_gap_hybrid}.
The single-mode system shows the expected exponential suppression of bit flips. 
Note that $\Gamma_{\rm bf}$ is larger then the corresponding dissipative counterpart, in agreement with Ref.~\cite{gautier_combined_2022}.
For the two-mode system, even in the most adiabatic case we observe large deviation with respect to the single mode bit-flip error rate. 
Moving away from the adiabatic regime, we observe that $\Gamma_{\rm bf}$ increases with the photon number, signaling that the cat qubit is completely degraded.
This analysis shows that, even in the absence of cross-Kerr nonlinearities, self-Kerr in the memory of a dissipative cat qubit has extremely detrimental effect.
This is in line with the prediction of the semiclassical analysis in Sec.~\ref{sec:semiclassical_analysis}.

\section{Details of the semiclassical analysis}
\label{app:semiclassical}

The semiclassical equation of motion for the $\alpha$ and $\beta$ $c$-numbers describing the memory and buffer are
\begin{equation}\label{eqs:semiclassical}
\begin{split}
&\dot\alpha =
-\frac{\kappa_a}{2}
\alpha
+ \rmi K_a|\alpha|^2\alpha
+ \rmi \chi|\beta|^2\alpha
- 2\rmi g_2\beta\alpha^\ast,
\\
&\dot\beta =
-\frac{\kappa_b}{2}\beta
+ \rmi K_b|\beta|^2\beta
+ \rmi \chi|\alpha|^2\beta
- \rmi g_2\,\alpha^2
- \rmi \varepsilon_b.
\end{split}
\end{equation}

\subsection{Adiabatic approximation}
\label{sec:adiabatic_approximation}

We now derive the conditions under which the buffer mode can be
adiabatically eliminated.
Assuming $\kappa_b$ is the dominant rate, the characteristic buffer timescale is
$
\tau_b \sim 1/{\kappa_b}.
$
To make this assumption, we have $|g_2 |\alpha|^2|, \,| \chi|\alpha|^2| \ll \kappa_b$
Adiabatic elimination requires that the memory variables change only
weakly over this interval,
\begin{equation}
|\dot{\alpha}| \tau_b \ll |\alpha|.
\end{equation}
Using Eq.~\eqref{eqs:semiclassical} gives the hierarchy
\begin{equation}
\kappa_b
\gg
\left\{
|g_2||\beta|,
\kappa_a,
|\Delta(t)|,
|K_a||\alpha|^2,
|\chi||\beta|^2
\right\}.
\label{Eq:condition_regularity}
\end{equation}
Under these conditions the buffer amplitude follows the memory
quasi-instantaneously.
Then, setting $\dot\beta\simeq0$ yields
\begin{equation}
\beta_{\rm ad}
=
\frac{\rmi(g_2\alpha^2+\varepsilon_b)}
{-\kappa_b/2+\rmi\chi|\alpha|^2}.
\label{eq:beta_ae}
\end{equation}

Notice that the adiabaticity condition for $\alpha$ to be ``slow'' needs to be satisfied for all the states explored by $\alpha$, even during a bit-flip event where $\alpha \to - \alpha$.
As the memory passes through the vacuum state $\alpha =0$, we get $|\beta| \simeq 2 \varepsilon_b/\kappa_b$.
In particular, using Eq.~\eqref{eqs:semiclassical}, the term
$\chi|\beta|^2$ still needs to be slow with respect to $\kappa_b$.
We the find 
\begin{equation}
 \chi|\beta|^2 \ll \kappa_b\quad\Rightarrow\quad 4\chi \varepsilon_b^2/\kappa_b^2 \ll \kappa_b.
\end{equation}

\subsection{Analysis of the stable point in the phase-locked regime}
\label{App:Stability_semiclassical}

The fixed points derived in Eq.~\eqref{eq:fixedpoints} exist provided $\vert \varepsilon_b \vert>\frac{\kappa_a\kappa_b}{8g_2}.$
Notice also the presence of the saddle point 
$x_{a,\mathrm{sad}}=0,~y_{b,\mathrm{sad}}=2\vert \varepsilon_b \vert/\kappa_b.$
Moreover, linearizing the equations of motion along any point $(x_a, 0, 0,y_b)$ on this plane, and considering small deviations $(\delta y_a, \delta x_b)$, we obtain the systen
\begin{equation}
\label{eq:jacobian}
\frac{\rmd}{\rmd t}
\begin{pmatrix}
\delta y_a \\
\delta x_b
\end{pmatrix}
=
\begin{pmatrix}
-\dfrac{\kappa_a}{2}-2g_2 y_b & -2g_2 x_a \\
2g_2 x_a & -\dfrac{\kappa_b}{2}
\end{pmatrix}
\begin{pmatrix}
\delta y_a \\
\delta x_b
\end{pmatrix}.
\end{equation}

In the limit of vanishingly small $\kappa_a$, the eigenvalues of this matrix are 
\begin{equation}
    \lambda_\perp^\pm
    =
    -\left(g_2 y_b+\frac{\kappa_b}{4}\right)
    \pm
    \sqrt{
    \left(\frac{\kappa_b}{4}-g_2 y_b\right)^2
    -4g_2^2 x_a^2
    }.
\end{equation}
As long as $y_b>0$, the real parts of the eigenvalues of this matrix remain negative and this manifold is locally attractive. However, this is no longer true in the domain $y_b<0$ and $|y_b|>\frac{\kappa_b}{4 g_2}$. The intuition behind is that if $y_b<0$, the memory is subject to a reverse squeezing drive that contracts $x_a$ and dilates $y_a$. 

Note that the condition $|y_b|>\kappa_b/4 g_2$ with $|y_b|\sim |y_{b,\mathrm{sad}}|=2\vert \varepsilon_b \vert/\kappa_b$, corresponds to the adiabatic condition required for the buffer to follow the memory (see Appendix \ref{sec:adiabatic_approximation}). In the non-adiabatic regime, the system is therefore able to amplify transverse components, an effect observed also in the quantum trajectory simulation in Fig.~\ref{fig:two_mode_nonadiabatic_traj}.

Finally, to obtain the estimate of the effect of $\chi$, let us define the local transverse locking gap as
\begin{equation}
    \Delta_\perp(x_a,y_b) = \min_\pm \operatorname |\lambda_\perp^\pm |.
\end{equation}
The effect of $s_{pert}$ remains small if the nonlinear frequency shifts are smaller than the transverse locking gap along in the explored region $(x_a,y_b)$:
\begin{equation}
\begin{split}
    |\Omega_a^0(x_a,y_b)|\ll \Delta_\perp(x_a,y_b),\\
    |\Omega_b^0(x_a,y_b)|\ll \Delta_\perp(x_a,y_b).
\end{split}
\end{equation}
Using the approximate switching-region bounds between the fixed points and the saddle point,
$|x_a|\lesssim x_{a,\mathrm{st}}$
and
$|y_b|\lesssim \frac{2\vert \varepsilon_b \vert}{\kappa_b}$,
a simple sufficient condition is
\begin{align}
|\chi|\left(\frac{2\vert \varepsilon_b\vert}{\kappa_b}\right)^2
\ll
\Delta_{\rm lock},\qquad 
|\chi|x_{a,\mathrm{st}}^2
\ll
\Delta_{\rm lock} ,
\end{align}
where
\begin{equation}
    \Delta_{\rm lock}
    \sim
    \min\left(
    \frac{4g_2^2\alpha^2}{\kappa_b},
    \frac{\kappa_b}{4}
    \right).
\end{equation}

As we are outside of the adiabatic regime, $\Delta_{\rm lock} \simeq {\kappa_b}/{4}$, retrieving the estimate in the main text.

\section{Spectral statistics of quantum trajectories}\label{sec:app_dqc}

\begin{figure}[t]
\centering
\includegraphics[width=0.48 \textwidth]{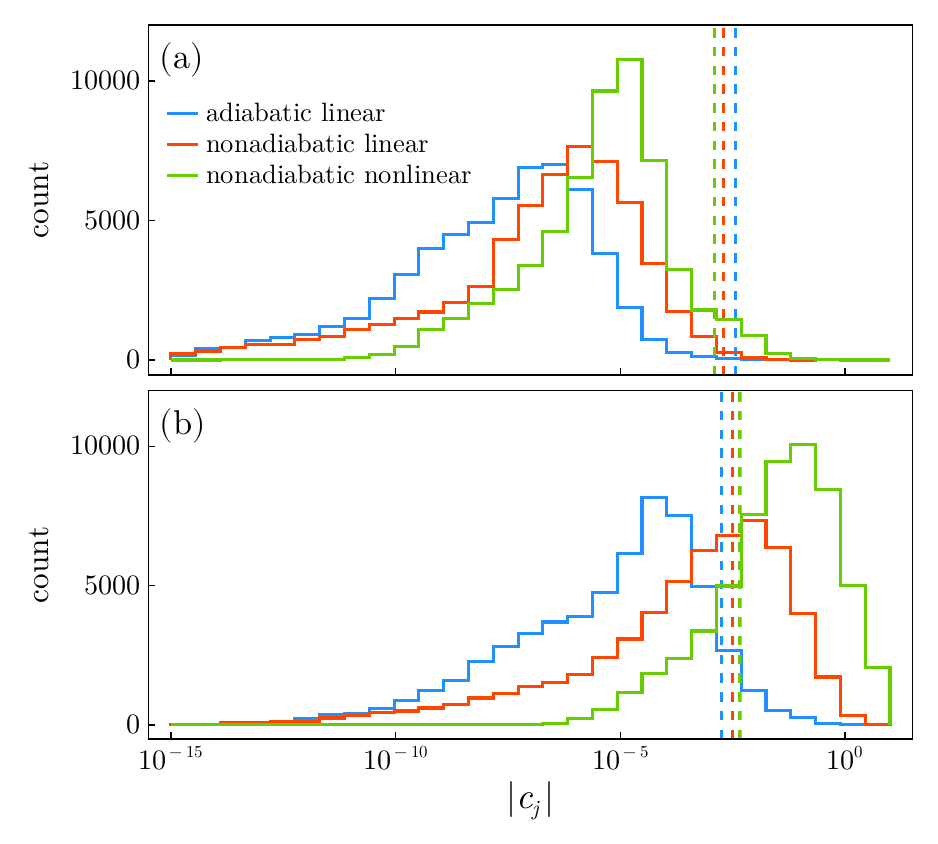}\vspace{0.02em}
\caption{Distributions of the spectral coefficients defined in Eq.~\eqref{eq:spectral_coefficients} and location of the cutoff $c_{\rm min}$ (dashed lines) for the quantum trajectories associated to the three models studied in Fig.~\ref{fig:ssqt}.
Panel (a) refers to the system's wave function between two bit-flip events.
Panel (b) refers instead to the system's wave function across a bit-flip events.
Blue curves indicates the adiabatic regime, red curves to the nonadiabatic linear regime, and green curves to the nonadiabatic nonlinear regime.
All parameters are fixed as in Fig.~\ref{fig:ssqt}.
}
\label{fig:ssqt_analysis}
\end{figure}

\begin{figure}[t]
\centering
\includegraphics[width=0.48 \textwidth]{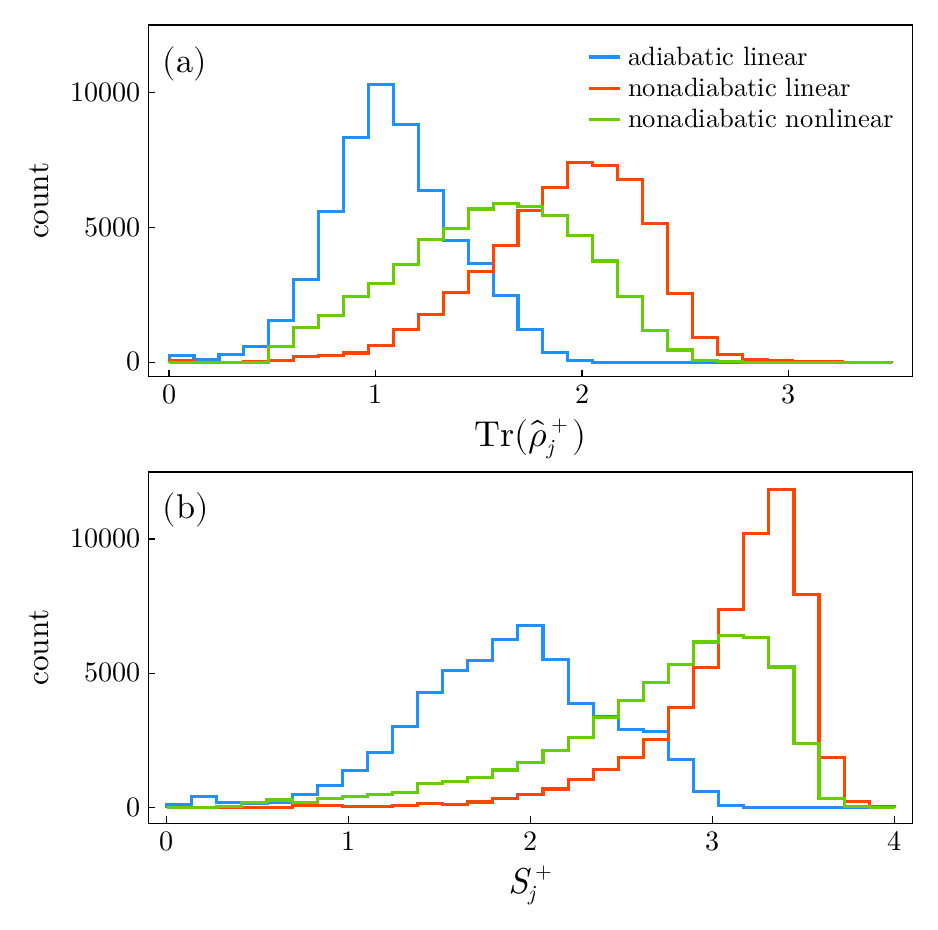}\vspace{0.02em}
\caption{
Histograms of (a) the traces of the unnormalized operators $\hat{\rho}_j^+$ defined in Eq~\eqref{eqs:decomposition} and (b) the Von Neumann entropies corresponding to the normalized $\hat{\rho}_j^+$, $S_j^+ = -\operatorname{Tr}\left(\hat{\rho}_j^+\,\log\,\hat{\rho}_j^+\right)$.
The histograms are realized for all the eigenstates of the Liouvillian.
Blue curves indicates the adiabatic regime, red curves to the nonadiabatic linear regime, and green curves to the nonadiabatic nonlinear regime.
All parameters are fixed as in Fig.~\ref{fig:ssqt}.
}
\label{fig:trace_entropy}
\end{figure}

The spectral statistics of quantum trajectories (SSQT) used in Sec.~\ref{sec:dissipative_quantum_chaos} and introduced in Ref.~\cite{ferrari_dissipative_2025} addresses the question of which Liouvillian eigenstates and eigenvalues are activated along the dynamics of single quantum trajectories.
This is carried out by constructing the density matrix of a quantum trajectory at time $t$, i.e., $\hat{\rho}(t) = \ketbra{\Psi(t)}$, and decomposing it over the basis of the Liouvillian eigenvectors given by Eq.~\eqref{eq:liouvillian_spectrum}, so that
\begin{equation}\label{eq:spectral_coefficients}
    \hat{\rho}(t) = \sum_j c_j(t)\hat{\eta}_j, \quad c_j(t) = \operatorname{Tr}\left[\hat{\rho}(t)\hat{\sigma}_j^\dagger\right].
\end{equation}
One then considers the Liouvillian eigenstates for which $|c_j|>c_{\rm min}$, where the cutoff is proportional to the center of mass of the $|c_j|$, 
\begin{equation}\label{eq:spectral_cutoff}
    c_{\rm min} = \left(\sum_j|c_j|^2/\sum_j|c_j|\right)\times 10^{-k},
\end{equation}
and $k$ is chosen such that the left tails of the distribution of the spectral coefficients is taken into account (tipycally, $k \in[2, 4]$ \cite{ferrari_dissipative_2025}; in this article, we set $k=2$).
Notice also that there is, in principle, a gauge ambiguity in this definition.
We remove it using a physically motivated normalization, as discussed in the following section.

This approach captures the amount of delocalization of the stochastic wave function across the Liouvillian spectrum (see also Ref.~\cite{richter2025localizationdelocalizationquantumtrajectories}) and gives access to the collection of processes that build up $\ket{\Psi(t)}$.
When a small portion of the Liouvillian spectrum (usually, the steady state and some low-lying eigenstates) is involved in the quantum trajectory dynamics, the system displays integrable features \cite{ferrari_dissipative_2025}, regardless of the \textit{unactivated} remaining part of the spectrum.
When many eigenstates contribute to $\ket{\Psi(t)}$, their features are then crucial in determining integrable or chaotic dynamics.

In Fig.~\ref{fig:ssqt_analysis} we plot the histograms of the spectral coefficients $|c_j|$ and the associated cutoff $c_{\rm min}$ associated to the spectra reported in Fig.~\ref{fig:ssqt} in the main text.
In particular, Fig~\ref{fig:ssqt} (a)  refers to the dynamics between two bit-flips and Fig~\ref{fig:ssqt} (b) to the dynamics across a bit-flip, while the color of the lines indicate the adiabatic linear (blue), the nonadiabatic linear (red) and the nonadiabatic nonlinear regime (green).
We observe that the peaks of the histograms in Fig~\ref{fig:ssqt} (a) are to the left of $c_{\rm min}$, and thus a few eigenstates are involved into the dynamics of $\hat{\rho}(t)=\ketbra{\Psi(t)}$.
In Fig.~\ref{fig:ssqt} (b) for the nonadiabatic linear and nonlinear regimes, instead, the peak of the histogram locates at the right of $c_{\rm min}$, reflecting the massive amount of eigenstates participating to the trajectory dynamics across the bit-flip.
It is also worth noting that all $c_{\rm min}$ are almost equivalent, confirming the effectiveness of Eq.~\eqref{eq:spectral_cutoff} as the spectral cutoff.\\

\subsection{Choice of the eigenstate normalization}
\label{App:normalization}

An important point of non-Hermitian matrices is that the bi-orthonormality condition fixes the normalization of $\hat{\eta}_j$ with respect to $\hat{\sigma}_j$, but there exists a gauge invariance such that $\operatorname{Tr}(\hat{\eta}_j^\dagger\hat{\eta}_j)$ is not fixed \textit{a priori}.
In all the calculations done in this paper, we impose $\operatorname{Tr}(\hat{\eta}_j^\dagger\hat{\eta}_j)=1$. This can be justified by the following argument. 
Given $\hat{\eta}_j$ with $\operatorname{Tr}(\hat{\eta}_j)=0$, we can construct the Hermitian, zero-trace, combination $\hat{\rho}_j = (\hat{\eta}^\dagger_j + \hat{\eta}_j)/2$ [or, equivalently, $\hat{\rho}_j =  \rmi(\hat{\eta}^\dagger_j - \hat{\eta}_j)/2$]. 
The $\hat{\rho}_j$ can be diagonalized to get
\begin{equation}
\begin{split}
    \hat{\rho}_j = \sum_{k=-N/2}^{N/2}p_k^{(j)}\ketbra{\Psi_k^{(j)}}, \quad \sum_{k=-N/2}^{N/2}p_k^{(j)}=0.
\end{split}
\end{equation}
We can then decompose $\hat{\rho}_j$ as the linear combination of two matrices, $\hat{\rho}_j = \hat{\rho}_j^+ - \hat{\rho}_j^-$,
\begin{equation}\label{eqs:decomposition}
\begin{split}
    \hat{\rho}^{\pm}_j = \pm\sum_{k=1}^{N/2}p_{\pm,\,k}^{(j)}\ketbra{\Psi_{\pm,\,k}^{(j)}},
\end{split}
\end{equation}
such that
\begin{equation}
    \sum_{k=1}^{N/2}p_{\pm,\,k}^{(j)}=\pm\mathcal{O}(1).
\end{equation}
The condition $\operatorname{Tr}(\hat{\eta}_j^\dagger\hat{\eta}_j)=1$ implies that $\hat{\eta}_j$ can be decomposed in linear combinations of objects with positive $\mathcal{O}(1)$-trace.
Furthermore, upon normalization, we can associate to $\hat{\eta}_j$ its entropy (or any other measure of delocalization) through the corresponding entropy of $\hat{\rho}_j^{\pm}$, $S_j^{\pm} = -\operatorname{Tr}\left(\hat{\rho}_j^{\pm}\,\log\,\hat{\rho}_j^{\pm}\right)$.
In Fig.~\ref{fig:trace_entropy} (a) we plot the histogram of the unnormalized $\operatorname{Tr}(\hat{\rho}_j^+)$ for the three models studied in Fig.~\ref{fig:ssqt}, showing indeed that $\operatorname{Tr}(\hat{\rho}_j^+)\sim\mathcal{O}(1)$.
In Fig.~\ref{fig:trace_entropy} (b) we plot the histograms of the entropies $S_j^+$ (upon the normalization of $\hat{\rho}_j^+$) associated to the right eigenvectors $\hat{\eta}_j$ again for the three models studied in Fig.~\ref{fig:ssqt} in the main text.

\end{document}